\documentclass[twocolumn]{aastex631}
\usepackage{amsmath}
\usepackage{graphicx}
\shorttitle{Analytics of MET Alignment and Disruption}
\shortauthors{Lazarian}
\begin{document}
\title{Towards Analytical Description of Dust Alignment and Disruption by Mechanical Torques}
\author{A. Lazarian}
\affiliation{Department of Astronomy, University of Wisconsin--Madison}
\begin{abstract}
We consider the alignment of irregular dust grains by mechanical torques (METs) that arise through dust-gas drift. We explore the possibility of analytically describing the statistical properties of METs to provide a reliable way to describe their alignment efficiency for different astrophysical settings. We employ a statistical ensemble for 180 generated compact convex irregular grains. In analogy to radiative torque (RATs) studies, we focus on the alignment of grains with stable high-$J$ attractors.
Unlike RATs, METs of individual demonstrate significant variations in the functional depencences of torques calculated in the grain coordinates. However, the torques averaged over the grain ensemble demonstrate similarity to RAT torques, opening a way for an analytical description of MET alignment. We consider MET alignment with and without magnetic dissipation for different values of grain drift through the thermal gas. Similarly to the case of RATs, magnetic dissipation, which can be paramagnetic for silicate grains and super-paramagnetic/ferromagnetic for grains with inclusions, increases the degree of grain alignment. We illustrate the METs effects considering grain alignment in a few astrophysical settings including dark clouds and circumstellar accretion disks. Ambipolar diffusion and turbulence can drive METs in the environments where RATs are not efficient. Our estimates of rotational disruption in turbulent interstellar medium demostrate that METs can significantly affect the interstellar dust size distribution. METs has a potential for explaining polarization features of circumstellar accretion disks. 
\end{abstract}

\section{Introduction}
The alignment of interstellar dust grains has been a long-standing problem in astrophysics since the discovery of interstellar starlight polarization by \citet{Hall1949} and \citet{Hiltner1949}. These observations established aligned nonspherical grains as tracers of interstellar magnetic fields, although their alignment mechanism remained uncertain for decades. The magnetic-alignment paradigm began with paramagnetic dissipation proposed by \citet{DavisGreenstein1951}. The efficiency of the mechanism was improved through improved magnetic responses, including the superparamagnetic-grain interpretation of \citet{JonesSpitzer1967}, as well as the suprathermal rotation and body-fixed torques discussed by \citet{Purcell1979}; see also \citet{SpitzerMcGlynn1979}. The range of possible mechanisms motivated more quantitative tests. 

A different approach emerged with radiative torques (RATs), initially proposed by \citet{DolginovMitrofanov1976} and established numerically for irregular grains by \citet{DraineWeingartner1996,DraineWeingartner1997}. A major step toward predictive theory was the analytical model (AMO) of helical-grain RATs developed by \citet{LH2007rat}, which reproduced key properties of numerically calculated torques. The combination of RAT spin-up and magnetic relaxation was subsequently explored by \citet{LazarianHoang2008} and \citet{HoangLazarian2016}. RAT alignment theory has been supported by a broad range of observations. The review in \citet{AnderssonLazarianVaillancourt2015} summarized observational and theoretical evidence. RAT theory has been accepted by the community as leading explanation of astrophysical grain alignment.

Mechanical alignment offers an alternative route, with grain--gas drift rather than radiation providing the anisotropic forcing. \citet{Gold1952} proposed stochastic alignment by directed gas supersonic bombardment of idealized elongated grains. \citet{Lazarian1994} developed a quantitative treatment of Gold-type alignment including internal grain relaxation. Subsequent work examined mechanical alignment under astrophysical conditions including drift associated with ambipolar diffusion \citep[e.g.,][]{Lazarian1995,RobergeHananyMessinger1995,DeschRoberge1997}. Applying the helical-grain insight developed for RATs, particularly AMO, \citet{LH2007met} introduced regular mechanical torques (METs) on irregular grains under gas bombardment. The paper concluded that METs can align grains even for subsonic grain-gas relative motions.  

Numerical studies of METs by \citet{DasWeingartner2016} and \citet{HoangChoLazarian2018} further demonstrated systematic alignment by grain--gas drift. \citet{HoangEtAl2022Disk} explicitly studied MET alignment in protostellar disks, including mechanical spin-up, internal relaxation, and alignment with either the magnetic field or the grain--gas drift. We compare their results with ours in Section~\ref{sec:disk-prior-work}. 

MET alignment remains less quantitatively characterized than RAT alignment. Numerical torque calculations have covered relatively few grain geometries and gas--surface prescriptions. The RAT analytical model does not yet have an equally reliable counterpart for MET torque functions, whose angular dependence varies with grain shape and collision physics. 

The present paper addresses the limitation above through direct numerical calculations for 180 independently generated compact irregular grains. We began with 30 grains to establish the numerical method, then added 150 shapes. They form one sample: the first 30 provide a direct check, while the full sample is used for the reported torque statistics and alignment factors. A key result is that magnetically enhanced MET alignment can remain strong at exactly perpendicular grain--neutral drift, making ambipolar diffusion a potentially important alignment mechanism rather than an intrinsically unfavorable case. In dark clouds where the radiation field is attenuated, MET alignment may reach grains for which RAT driving is inefficient. The relevant size range, however, is determined jointly by the grain--neutral drift, suprathermal spin-up, and magnetic relaxation. Our work is inspired by the study in \citet{DasWeingartner2016} where the convex shapes were generated and the METs were explored. Using this general approach, we calculate a larger shape ensemble, test the resulting attractor stability, and explore astrophysical implications.  

\cite{LH2007met} demonstrated that grains can align in both positions corresponding to high angular momentum, i.e., high-J attractor points and low angular momentum, i.e. low-J attractor points. Throughout this paper we calculate alignment exclusively from stable high-$J$ states, assuming perfect internal alignment and equal statistical weights for the grain shapes. We assign $R=0$ to grains lacking such a state and do not calculate alignment at low-$J$ attractors, as the grains at this state rotate subthermally and lose internal alignment \citet{LH2007rat}.

For sustained ambipolar drift and disk conditions, we adopt the explicit long-time approximation that gas impacts can move grains into high-$J$ attractors when they exist. For shocks, the finite grain--gas interaction time must be compared with the calculated relaxation time, and the saturated factor need not be reached. 

Section~2 defines the grain model and magnetic response. Section~3 develops the collision torques and tests their ensemble-averaged angular dependence against AMO. Section~4 derives the precession averages and high-$J$ stability criteria. Sections~5 and 6 present the alignment factors and their magnetic enhancement. Section~7 compares METs and RATs. Sections~8 and 9 examine shocks, ambipolar drift, circumstellar flows, and disks. Section~10 evaluates mechanical disruption. The Discussion and Summary follow in Sections~11 and 12.

\section{Grain model and magnetic relaxation}

\begin{figure}[t]
\centering
\includegraphics[width=\columnwidth]{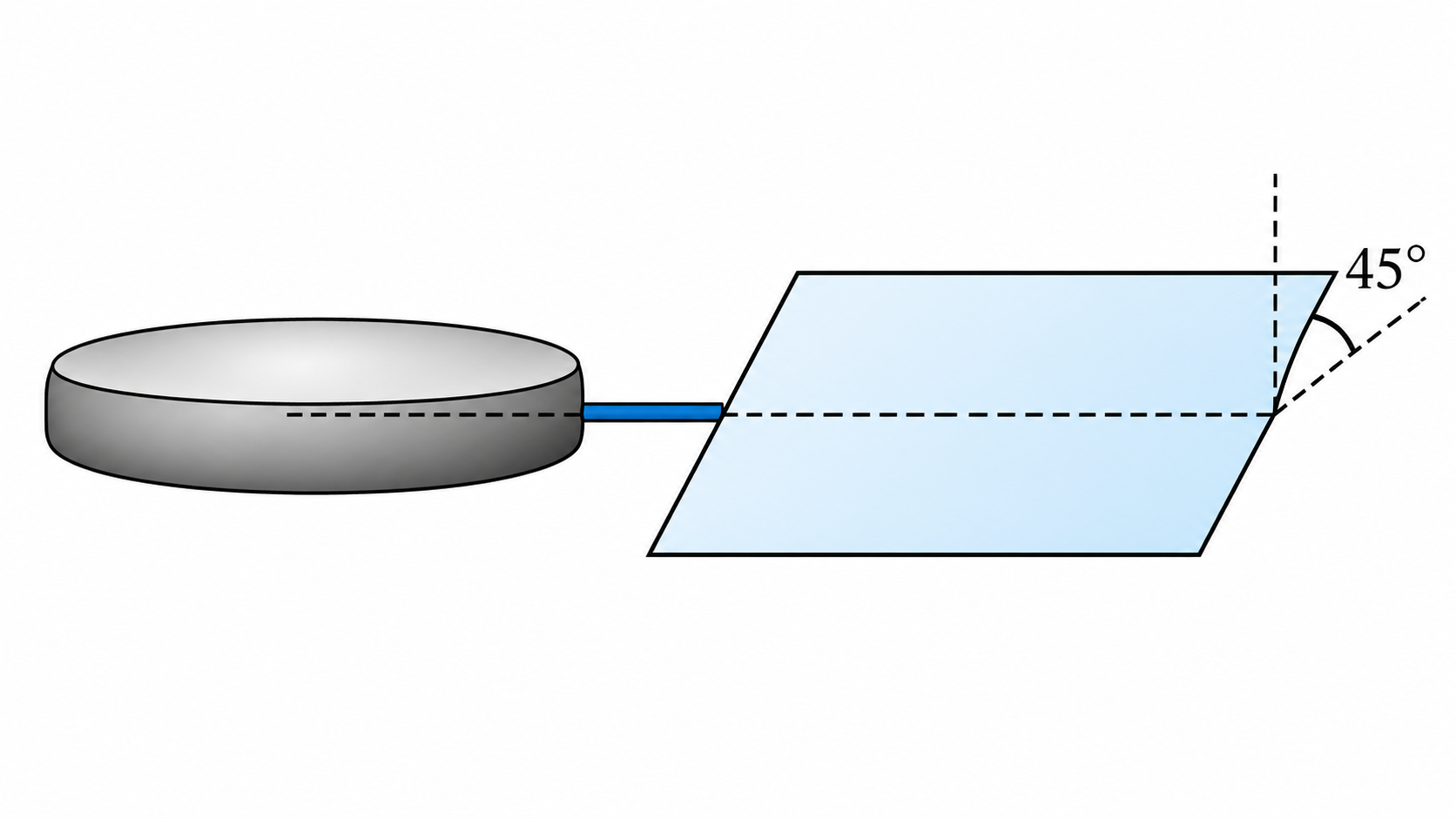}
\caption{Schematic of the analytical model of radiative torques (AMO) introduced by \citet{LH2007rat}. An inclined mirror is attached to an oblate grain to represent helicity. In the standard version, the mirror is tilted by $45^\circ$ relative to the grain body. The dashed horizontal line indicates the radiation direction. This simple geometry reproduces the basic angular behavior of the RAT components and provides the reference used below when we compare MET angular profiles with AMO.}
\label{fig:amo-schematic}
\end{figure}

In the standard AMO, grain helicity is represented by an inclined mirror attached to an oblate body (Figure~\ref{fig:amo-schematic}). The mirror provides the helicity-dependent RAT components $Q_{e1}$ and $Q_{e2}$, while the nonhelical body can contribute a precessional $Q_{e3}$. Thus radiative precession does not require grain helicity. This geometry provides a reference for our comparison with METs.

\subsection{Grain interaction with gas}
Throughout this paper, the fiducial gas--surface collision law assigns half of incoming atoms to specular reflection and half to sticking followed by normal re-emission. We apply this mixture to the \emph{mean} MET used for the torque statistics, attractor calculations, and spin-up estimates. The limiting collision laws are compared separately. The mixture is a prescribed surface model, rather than a thermodynamic equilibrium boundary condition. In particular, the fluctuations of individual impacts need not be isotropic or equal to the equilibrium gas-noise tensor.

We denote the moment of inertia about the axis of maximum inertia by $I_1$ and use 
\begin{equation}
J_{\rm th}=(I_1k_BT_g)^{1/2}
\end{equation}
for the thermal angular momentum. For size estimates, a spherical reference grain of radius $a$ and material density $\rho$ has $I_1=(8\pi/15)\rho a^5$ and $V/I_1=5/(2\rho a^2)$. We denote the rotational gas-damping time by $\tau_{\rm drag}$ and use the standard estimate 
\begin{equation}
\tau_{\rm drag}\simeq0.7\rho a/(n_gm_gv_{\rm th}), 
\end{equation}
where $n_g$, $m_g$, and $v_{\rm th}$ are the gas number density, particle mass, and thermal speed \citep{HoangLazarian2016,Hoang2020DisruptionReview}. The spherical coefficient fixes the reference gas-damping normalization. Actual drag tensors and gas-impact covariances depend on grain shape and the gas--surface interaction. The numerical estimates below use this stated normalization.

\subsection{Magnetic relaxation in different environments}\label{sec:delta-environments}
Magnetic dissipation changes the angular stability of the high-$J$ states. We measure its strength relative to gas damping by
\begin{equation}
 \delta_m(\omega)\equiv\frac{\tau_{\rm drag}}{\tau_{\rm mag}}
 =\frac{\tau_{\rm drag}VB^2K(\omega)}{I_1},\qquad
 K(\omega)=\frac{\chi''(\omega)}{\omega},
 \label{eq:delta-general}
\end{equation}
where $\chi''$ is the dissipative part of the grain magnetic susceptibility \citep{JonesSpitzer1967,LazarianHoang2008,HoangLazarian2016}. Here $n_g$ and $m_g$ are the gas-particle number density and mass, and $v_{\rm th}=(2k_{\rm B}T_g/m_g)^{1/2}$ is the thermal speed. Inserting the grain inertia and gas-damping estimates defined above gives, for a compact grain,
\begin{equation}
 \delta_m\simeq\frac{1.75B^2K(\omega)}{n_gm_gv_{\rm th}a}.
 \label{eq:delta-gas-scaling}
\end{equation}
Thus magnetic relaxation becomes stronger in larger magnetic fields and more rarefied gas. At fixed gas properties and fixed $K$, $\delta_m\propto a^{-1}$. The rotation dependence of $K(\omega)$ is treated in \citet{Lazarian2026FreqMag}.

For ordinary paramagnetic grains, the magnetic response is typically of order $K\sim10^{-13}\,$s at the low frequencies relevant to these estimates. The following examples show how the same grain can occupy different relaxation regimes:
\begin{itemize}
\item In the cold neutral ISM ($n_g=30\,{\rm cm^{-3}}$, $T_g=100\,$K, $B=5\,\mu$G), a $0.1\,\mu$m grain has $\delta_m\simeq0.07$ for $K=10^{-13}\,$s. The magnetic relaxation is weak.
\item In a dense molecular cloud ($n_{\rm H_2}=10^4\,{\rm cm^{-3}}$, $T_g=10\,$K, $B=30\,\mu$G), the same grain has $\delta_m\simeq0.02$. A $0.03\,\mu$m grain instead has $\delta_m\simeq0.05$. The cloud calculations below combine this magnetic response with the grain--neutral drift.
\item In a strongly magnetized disk surface at 30 au, the parameters of Section~\ref{sec:disk-drift-estimate} at $z=4H_g$ imply $n_g\simeq1.3\times10^7\,{\rm cm^{-3}}$ and $B=11.5\,$mG. For $a=0.1\,\mu$m and $K=6\times10^{-14}\,$s, $\delta_m\simeq0.6$. Such a field can make ordinary paramagnetic relaxation important despite the relatively high density.
\item Circumstellar winds span a wide range of field strengths and densities. For an illustrative outflow with $n_g=10^6\,{\rm cm^{-3}}$, $T_g=10^3\,$K, and $B=1\,$mG, a $0.1\,\mu$m grain with $K=10^{-13}\,$s has $\delta_m\simeq0.02$. The scaling in Eq.~\eqref{eq:delta-gas-scaling} determines the increase in relaxation farther into a magnetically dominated or more dilute outflow.
\end{itemize}
The numerical examples use atomic hydrogen in the cold neutral ISM and molecular gas in the other environments. They illustrate weak and moderate relaxation without changing the grain composition.

Magnetic inclusions can increase both the dissipative response $K(\omega)$ and the static susceptibility $\chi(0)$ \citep{JonesSpitzer1967,LazarianHoang2008,HoangLazarian2016}. Increasing $K$ raises $\delta_m$ and can stabilize high-$J$ alignment. Increasing $\chi(0)$ accelerates Larmor precession and favors magnetic-field rather than drift-axis alignment. Values $\delta_m\gtrsim1$--2 are therefore accessible in sufficiently magnetically enhanced grains. The size and frequency dependence of the inclusion response, including intermediate-$J$ states, is discussed by \citet{Lazarian2026FreqMag}. In the present MET calculations $\delta_m$ is evaluated at a fixed adopted $K$ for each example.

\subsection{Generation of Grain Shapes}

We construct a sample of $N=180$ irregular grain shapes by perturbing a spherical surface with random spherical-harmonic modes. The method is motivated by the Gaussian-random-sphere approach of \citet{DasWeingartner2016}, although our grains are convex polyhedra rather than the nonconvex surfaces studied there. In particular, we represent the initial radial surface schematically as
\begin{equation}
 r(\theta,\phi)=r_0\exp\!\left[\sum_{\ell=2}^{6}\sum_{m=-\ell}^{\ell}c_{\ell m}Y^{\rm(real)}_{\ell m}(\theta,\phi)\right],
\end{equation}
where the $Y^{\rm(real)}_{\ell m}$ are real spherical harmonics and the random coefficients have an amplitude weighting proportional to $\ell^{-\alpha/2}$. The monopole is absorbed into the radius normalization, while the dipole is excluded from the imposed shape perturbations. We use two angular spectral indices, $\alpha=2$ and 3, and normalize the rms logarithmic radial perturbation to 0.25. The $\alpha=2$ shapes contain more small-scale structure.

For each realization, the perturbed radial surface is sampled in three dimensions and its convex hull is formed. The convex hull produces a closed polyhedron without cavities or shadowed facets. We therefore use single-impact gas collisions and exclude multiple collisions within a grain. 

\begin{figure*}[t]
\centering\includegraphics[width=.98\textwidth]{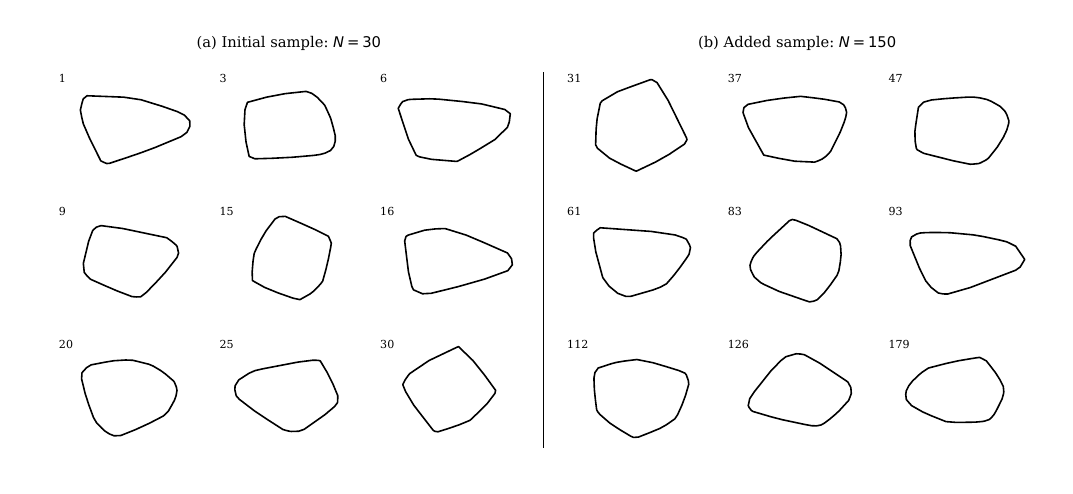}
\caption{Projected silhouettes of grains from the sample. \textit{Left:} nine of the first 30 shapes, used to establish and check the high-$J$ calculation. \textit{Right:} nine of the 150 additional shapes included in the final alignment and torque calculations. Numbers identify individual grain realizations. All shapes use the same effective-volume normalization.}
\label{fig:shapes}
\end{figure*}

The sample contains 90 realizations for each $\alpha$, including 15 of each type in the initial 30-grain test. Figure~\ref{fig:shapes} shows representative members of the initial and added samples in separate panels. Each polyhedron is rescaled to volume $V=4\pi a^3/3$, defining the effective radius $a=(3V/4\pi)^{1/3}$. We calculate its center of mass and moments of inertia for uniform material density and use the axis of largest moment $I_1$ as the reference axis. Across the sample, sphericity ranges from $0.87$ to $0.94$ and $I_1/I_3$ from $1.1$ to $3.0$. 

All shapes use seed 20261106, 420 near-equal-area surface directions per realization, and the same convex-hull and principal-inertia-axis construction. The surface sampling follows the public \href{https://github.com/budizhao/SHPSG}{SHPSG} generator. The first 30 grains reproduce the earlier torque results at $s_d=0.3$, 1, and 3. For the full ensemble we evaluate 13 drift strengths between $s_d=0.1$ and $5$. High-$J$ states and Rayleigh factors are calculated at $s_d=0.3$, 1, and 3. The other ten drift speeds characterize the torque functions.

\section{Collision torque}\label{sec:geometry}

\subsection{Torques in grain and laboratory frames}

We consider gas bombardment in the presence of relative grain--gas motion, described by a drift velocity $v_{\rm drift}$. Such drift can arise from turbulent grain acceleration in MHD dust--gas simulations \citep{MoseleyTeyssierDraine2023,MoseleyTeyssier2025}. 
For a shifted Maxwellian gas, with drift parameter $s_d=v_{\rm drift}/v_{\rm th}$, we write
\begin{equation}
 \boldsymbol\Gamma_{\rm MET}=\Gamma_0\mathbf Q,\qquad
 \Gamma_0=n_gm_gv_{\rm th}^2a^3.
\end{equation}
The torque exerted on each surface facet follows from the flux of incoming gas particles and the momentum they deliver. Let $\mathbf s=\mathbf v_g/v_{\rm th}$ be the gas velocity relative to the grain in thermal-speed units, with shifted Maxwellian $f(\mathbf s)=\pi^{-3/2}\exp[-|\mathbf s-\boldsymbol\mu|^2]$ and mean $\boldsymbol\mu=-s_d\hat{\mathbf v}$. For a facet with outward normal $\hat{\mathbf n}$, only particles with $s_n\equiv\mathbf s\cdot\hat{\mathbf n}<0$ reach the surface. The incident number flux per unit area and the corresponding momentum flux delivered to an absorbing surface follow from the surface-impact formulation used in mechanical-torque calculations (e.g., \citealt{DasWeingartner2016,HoangChoLazarian2018}):
\begin{align}
 \mathcal F_{\rm in}&=n_gv_{\rm th}\!\int_{s_n<0}(-s_n)f(\mathbf s)\,d^3s,
 \label{eq:met-number-flux}\\
 \boldsymbol\Pi_{\rm in}&=n_gm_gv_{\rm th}^2\!\int_{s_n<0}(-s_n)\mathbf s\,f(\mathbf s)\,d^3s.
 \label{eq:met-momentum-flux}
\end{align}
The first quantity gives the impact rate. The second gives the incident momentum transfer before reflection or re-emission.

For the shifted Maxwellian, $s_n$ is normally distributed with mean $\mu_n=-s_d\hat{\mathbf v}\cdot\hat{\mathbf n}$ and dispersion $\sigma=1/\sqrt{2}$. Evaluation of these collision fluxes reduces to truncated Gaussian moments. The flux integrals involve the two moments $M_k\equiv\int_{-\infty}^{0}s_n^k p_n(s_n)\,ds_n$ ($k=1,2$), where $p_n$ is the normal-velocity probability density. Their analytical values are
\begin{align}
 M_1&=\mu_n\Phi(-x)-\sigma\varphi(x),\\
 M_2&=(\mu_n^2+\sigma^2)\Phi(-x)-\mu_n\sigma\varphi(x),
 \qquad x=\mu_n/\sigma,
\end{align}
where $\varphi(x)=(2\pi)^{-1/2}\exp(-x^2/2)$ and $\Phi(x)=\int_{-\infty}^{x}\varphi(t)\,dt$ are the standard normal density and cumulative distribution. In particular, the incident number flux is $\mathcal F_{\rm in}=-n_gv_{\rm th}M_1$.

In units of $n_gm_gv_{\rm th}^2$, the incident momentum flux is $\mathbf P_{\rm arr}=(-M_2+\mu_nM_1)\hat{\mathbf n}-M_1\boldsymbol\mu$. We integrate the gas velocity distribution analytically rather than simulating individual impacts \citep[cf.][]{HoangChoLazarian2018}. Applying the surface momentum-transfer prescriptions of \citet{DasWeingartner2016,HoangChoLazarian2018}, specular reflection gives $\mathbf P_{\rm spec}=-2M_2\hat{\mathbf n}$. For sticking followed by normal re-emission at the prescribed outward speed $0.6v_{\rm th}$, the momentum transfer is $\mathbf P_{\rm stick}=\mathbf P_{\rm arr}+0.6M_1\hat{\mathbf n}$. The torque from each surface facet is obtained from the cross product of its lever arm about the grain center of mass with the transferred momentum: $d\mathbf Q=(dA/a^2)[(\mathbf r-\mathbf r_{\rm cm})/a]\times\mathbf P$. The model excludes multiple surface collisions and detailed energy exchange.

These equations give the mean collision torque. To distinguish it from random impact excitation, let $\Delta\mathbf J(\mathbf r,\mathbf s,\alpha)$ be the angular-momentum impulse of a single impact at surface position $\mathbf r$, incident velocity $\mathbf s$, and collision outcome $\alpha$. For independent impacts, the raw second jump moment gives the local covariance per unit time,
\begin{equation}
\begin{split}
2D_{ij}^{\rm coll}={}&\sum_\alpha\int_{\partial V}\!dA
\int_{s_n<0}\!d^3s\,n_gv_{\rm th}(-s_n)f(\mathbf s)\\
&\quad\times P_\alpha\,\Delta J_i\Delta J_j.
\end{split}
\label{eq:met-shot-noise}
\end{equation}
Here $P_\alpha$ is the prescribed probability of the collision outcome, and re-emission fluctuations must be included in $\Delta\mathbf J$ when present. The mean torque instead uses the first jump moment. The first impact moment gives the systematic torque, while the second gives random rotational excitation. We calculate only the mean torques and do not predict stochastic capture rates.

More generally, for a fraction $f_{\rm sp}$ of specular outcomes, the mixed impulse is
\begin{equation}
 \mathbf P_{\rm mix}=(1-f_{\rm sp})\mathbf P_{\rm stick}
                      +f_{\rm sp}\mathbf P_{\rm spec},
 \qquad 0\le f_{\rm sp}\le 1.
 \label{eq:met-mixed-impulse}
\end{equation}
Each mixed-law torque component therefore obeys
$Q_{ei}^{\rm mix}=(1-f_{\rm sp})Q_{ei}^{\rm stick}+f_{\rm sp}Q_{ei}^{\rm spec}$.
We combine the torques before finding equilibria because stability and alignment depend nonlinearly on $f_{\rm sp}$. 

\begin{figure}[t]
\centering
\includegraphics[width=\columnwidth]{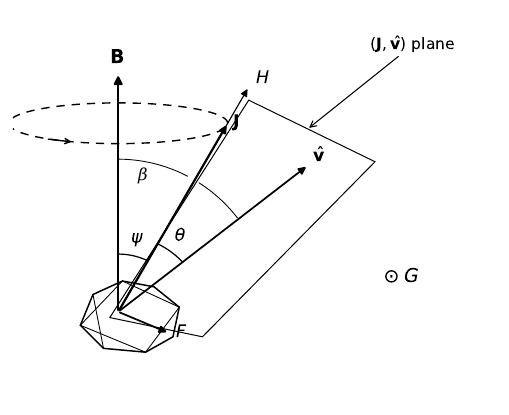}
\caption{Mechanical-torque geometry. The angles $\psi$, $\theta$, and $\beta$ separate $\mathbf J$, $\mathbf B$, and drift $\hat{\mathbf v}$. Dashed: precession about $\mathbf B$. Torques $H$, $F$, and $G$ respectively change spin, grain--drift inclination, and azimuth. The axes need not be coplanar.}
\label{fig:hfggeometry}
\end{figure}

Following the scattering-frame description of RATs in \citet{LH2007rat}, we account for perfect internal relaxation and describe the MET torques by $Q_{e1}$, $Q_{e2}$, and $Q_{e3}$. The torques are defined relative to the grain's maximum-inertia axis $\hat{\mathbf u}$ and the directed grain--gas drift $\hat{\mathbf v}$. For $\Theta=\cos^{-1}(\hat{\mathbf u}\cdot\hat{\mathbf v})$ (the angle $\theta$ in Figure~\ref{fig:hfggeometry}), 
\begin{equation}
 \hat{\mathbf e}_1=\hat{\mathbf v},\qquad
 \hat{\mathbf e}_2=\frac{\hat{\mathbf u}-\cos\Theta\,\hat{\mathbf v}}{\sin\Theta},\qquad
 \hat{\mathbf e}_3=\hat{\mathbf e}_1\times\hat{\mathbf e}_2.
 \label{eq:met-scattering-basis}
\end{equation}
 After averaging over rapid grain rotation about $\hat{\mathbf u}$, the components of the dimensionless collision torque $\overline{\mathbf Q}$ are
\begin{equation}
 Q_{ei}(\Theta;s_d)=\overline{\mathbf Q}(\Theta;s_d)\cdot\hat{\mathbf e}_i,
 \qquad i=1,2,3.
 \label{eq:met-Qe-first}
\end{equation}
As in LH07, these are \emph{scattering-frame} components defined using the grain orientation. We use them to compare individual grains, ensemble means, and the AMO torque shapes.

The torque components relevant for rotational dynamics follow by projecting these $Q_{ei}$ onto the grain spin direction, the direction of increasing grain--drift angle, and the azimuthal direction:
\begin{align}
 H(\Theta)&=Q_{e1}(\Theta)\cos\Theta+Q_{e2}(\Theta)\sin\Theta,\\
 F(\Theta)&=-Q_{e1}(\Theta)\sin\Theta+Q_{e2}(\Theta)\cos\Theta,\\
 G(\Theta)&=Q_{e3}(\Theta),
 \label{eq:met-Qe-projection}
\end{align}
where $H$ produces spin-up or spin-down, $F$ changes the angle $\Theta$ between $\hat{\mathbf u}$ and the drift, and $G$ produces precession about the drift direction. 

Figure~\ref{fig:hfggeometry} shows these dynamical directions. With angular momentum $j$ signed along the maximum-inertia axis, $\dot\Theta=\Gamma_0F/j$, so the sign of $j$ matters for angular stability. The angle $\Theta$ is measured from the drift, whereas $\psi$ is the angle from the grain axis to the magnetic field (for $j<0$, the angle between $\mathbf J$ and $\mathbf B$ is $\pi-\psi$). In the rapid-Larmor-precession regime, the evolution of $\psi$ is determined by the precession-averaged torque $F_B(\psi)$. The angle $\beta$ is between drift and magnetic field.

For the statistical comparison, the mean torque component for a set of $N$ grain shapes is
\begin{equation}
\langle Q_{ei}(\Theta)\rangle_N
=\frac{1}{N}\sum_{j=1}^{N}Q_{ei}^{(j)}(\Theta).
\label{eq:met-ensemble-mean}
\end{equation}
For an individual grain $j$, we define the ratio of the peak amplitudes of the two torque components as
\begin{equation}
q_{\max}^{(j)}
=\frac{\displaystyle\max_{\Theta}\bigl|Q_{e1}^{(j)}(\Theta)\bigr|}
{\displaystyle\max_{\Theta}\bigl|Q_{e2}^{(j)}(\Theta)\bigr|}.
\label{eq:met-qmax}
\end{equation}
This ratio measures the relative amplitudes of the spin-up and orienting torque components for each grain, following the convention introduced for RATs by \citet{LH2007rat}.
Grains with opposite helicities have different sign of $Q_{e1}$ and $Q_{e2}$ as demonstrated in \cite{LH2007rat}. For comparisons of \emph{angular dependences} alone, we reverse the common sign of $Q_{e1}$ and $Q_{e2}$ for grains with $Q_{e1}(0)<0$ before averaging. The component $Q_{e3}$ retains its original geometric sign, and $q_{\max}^{(j)}$ uses unsigned peak magnitudes. The familiar approximate AMO functions $(5\cos^2\Theta-2)/3$ and $\sin 2\Theta$ describe the torques for the standard $45^\circ$ mirror \citep{LH2007rat}. The complete spin-averaged AMO functions also depend on mirror inclination, which we compare with our MET results in Section~\ref{sec:amo-mirror}.

\begin{figure*}[t]
\centering
\includegraphics[width=.96\textwidth]{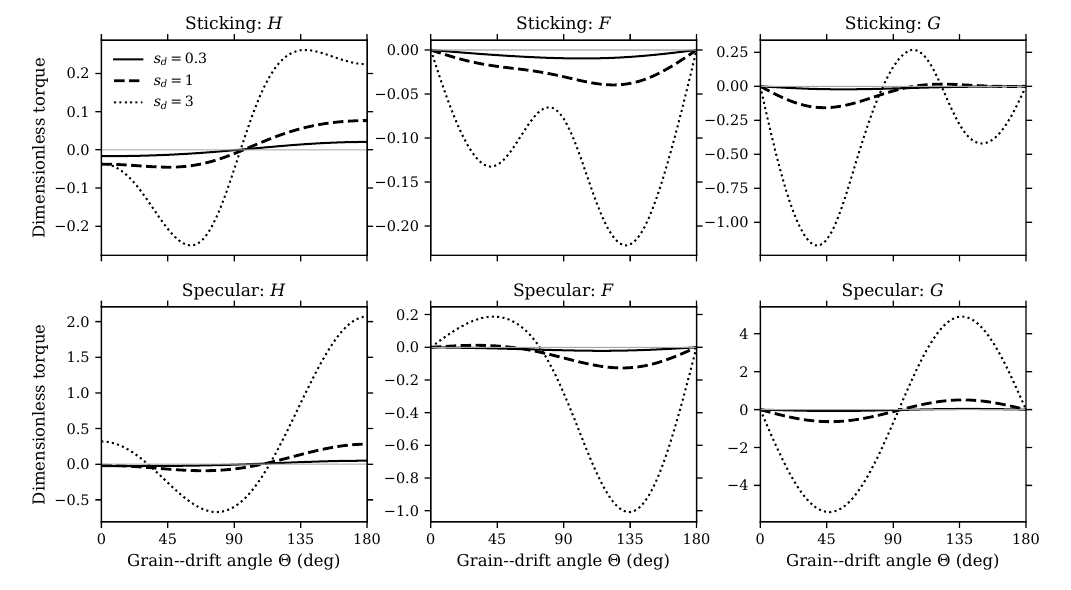}
\caption{Spin-averaged $H,F,G$ torque components of reconstructed grain 1 under the two limiting collision prescriptions, at three drift strengths. The default mixed-law functions are the pointwise means of these curves.}
\label{fig:torques}
\end{figure*}

\subsection{Statistical convergence and analytical torque representation for an ensemble of grains}
\label{sec:torque180}

All shapes have equal statistical weight. 
For an individual grain, stronger drift changes both torque amplitude and angular dependence. Averaging over the sample reduces the differences between individual torque curves and reveals an AMO-like mean (Figures~\ref{fig:torques} and \ref{fig:torques180}). Figure~\ref{fig:torquefit180} shows how the normalized mean evolves with $s_d$.

Figure~\ref{fig:torques180} compares the mean and grain-to-grain scatter for the initial 30 shapes and the full sample of 180. The median $q_{\max}$ changes little with sample size. For the full sample, it is about 1.4 at $s_d=0.3$ and 1.1 at $s_d=3$. The individual torque functions exhibit much greater scatter.

Figure~\ref{fig:qmax180} shows that the median $q_{\max}$ decreases over the sampled finite-drift range and converges with sample size. The error bars indicate how the median varies when the grain shapes are resampled. They do not represent random gas-impact fluctuations. For any individual grain with nonzero linear response $D_\parallel$, Eq.~\eqref{eq:met-Qe-weak} gives $q_{\max}\to4/3$ as $s_d\to0$. The sample median is fitted by
\begin{equation}
 q_{\max,{\rm med}}(s_d)\simeq \frac43+
 \frac{0.13s_d-0.27(s_d/1.16)^2}{1+(s_d/1.16)^2},
 \qquad 0\leq s_d\lesssim5.
 \label{eq:qmax-fit}
\end{equation}
This interpolation reproduces the calculated trend and the weak-drift limit $4/3$.  

We checked the weak-drift prediction independently with a \emph{second} sample of the same size, generated independently. Unlike the primary spherical-harmonic ensemble, these test shapes were built from 64 random surface directions per grain, perturbed by a random quadratic deformation and radial noise, and then convexified. Re-evaluating the facet impulses with the default collision law and averaging over 24 spin phases gave a median $q_{\max}\simeq1.3$ for $s_d=0.03$--0.3. The approach toward $4/3$ is therefore recovered for a different family of irregular shapes. This second sample tests only the weak-drift torque shape, not high-$J$ alignment.

The compact angular representation we find useful is, with $c=\cos\Theta$ and Legendre polynomials $P_\ell(c)$,
\begin{align}
\overline Q_{e1}(\Theta;s_d)&=A_2(s_d)P_2(c)+A_3(s_d)P_3(c)\\
            &\quad +A_4(s_d)P_4(c),\\
\overline Q_{e2}(\Theta;s_d)&=\sin\Theta\,[B_0(s_d)\\
              &\quad +B_1(s_d)P_1(c)+B_2(s_d)P_2(c)\\
              &\quad +B_3(s_d)P_3(c)],\\
\overline Q_{e3}(\Theta;s_d)&=C_1(s_d)\sin\Theta P_1(c).
\label{eq:met180-angular}
\end{align}
The coefficients $A_i(s_d)$, $B_i(s_d)$ and $C_1(s_d)$ describe the separately calculated ensemble means at each drift speed.  In the low-drift limit the leading terms give $\overline Q_{e1}\propto s_dP_2(c)$ and $\overline Q_{e2}\propto s_d\sin 2\Theta$. The numerical curves for increasing $s_d$ (Figure~\ref{fig:torquefit180}) show that the lowest-order angular structure remains recognizable while its relative amplitudes and higher-order contributions change. The $Q_{e3}$ curves appear almost unchanged with $s_d$ because each curve is divided by its own maximum in Figure~\ref{fig:torquefit180}. Their actual amplitude grows strongly: the peak magnitude of the ensemble mean is about 0.03 at $s_d=0.3$, 0.3 at $s_d=1$, and 2.6 at $s_d=3$ in our torque units. The mean precessional torque approximately follows $-\sin 2\Theta$. Individual grains retain additional angular structure that can affect stability, particularly near transverse drift. We therefore calculate their fixed points from individual torque functions.

The strength of $Q_{e3}$ has a geometric origin. Gas atoms transfer more momentum to some exposed facets than to others, producing a pressure torque on a nonspherical grain even in the absence of helicity. For a fore--aft symmetric grain, the leading spin-averaged nonhelical contribution has the form
\begin{equation}
 \boldsymbol\Gamma_{\rm geom}\propto
 (\hat{\mathbf u}\cdot\hat{\mathbf v})
 (\hat{\mathbf v}\times\hat{\mathbf u}),
 \label{eq:met-nonhelical-precession}
\end{equation}
which is perpendicular to the grain--drift plane and gives $Q_{e3}\propto\sin 2\Theta$. For a rapidly rotating grain with $\mathbf J\parallel\hat{\mathbf u}$, this torque changes the azimuth of $\mathbf J$ about the drift rather than spinning it up. In the idealized helical-grain picture, the in-plane components $Q_{e1}$ and $Q_{e2}$ depend more strongly on handedness, while a nonhelical body can supply $Q_{e3}$ \citep{LH2007met,HoangEtAl2022Disk}. The same mechanism exists for RATs: illumination of a nonhelical spheroid can produce $Q_{e3}$ even when its spin-up and polar aligning components vanish \citep{LH2007rat}. A large precessional component is therefore not a property unique to METs.

For our collision model, the signed sample-mean precessional torque is approximately $\overline Q_{e3}\simeq-0.3s_d^2\sin 2\Theta$ over $0.1\leq s_d\lesssim3$. This nearly quadratic dependence reflects the anisotropic momentum flux of the gas. The weak-drift sample-mean spin-up and polar components are instead linear in $s_d$ to leading order. Hence the characteristic ratio of the nonhelical precessional torque to the in-plane torques initially increases with $s_d$. The fitted coefficient describes the common geometric contribution of the sample, not every individual grain's angular function.

The growing relative contribution of $Q_{e3}$ results from changing the incident gas velocity distribution, not merely increasing the number of collisions. At fixed $s_d$ and $T_g$, increasing gas density multiplies all MET components equally. Likewise, increasing radiation intensity at fixed spectrum multiplies all RAT components equally. MET ratios change with drift strength, while RAT ratios can change with radiation wavelength relative to grain size \citep{LH2007rat}.

\begin{figure}[t]
\centering
\includegraphics[width=\columnwidth]{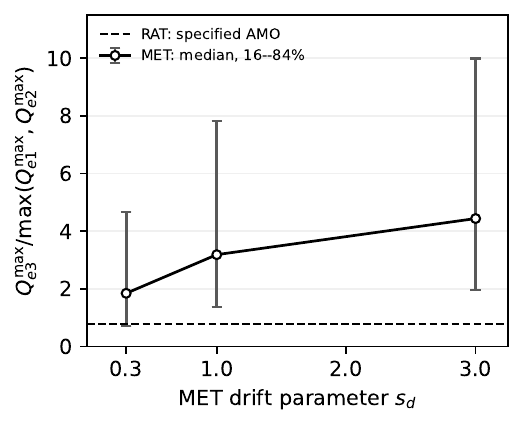}
\caption{Ratio of precessional to in-plane torque amplitudes. The MET curve shows the median for individual grains, with 16th--84th percentile bars. The dashed line shows the specified RAT AMO normalization from \citet{LH2007rat}.}
\label{fig:met-rat-q3-ratio}
\end{figure}

\begin{figure*}[t]
\centering\includegraphics[width=.98\textwidth]{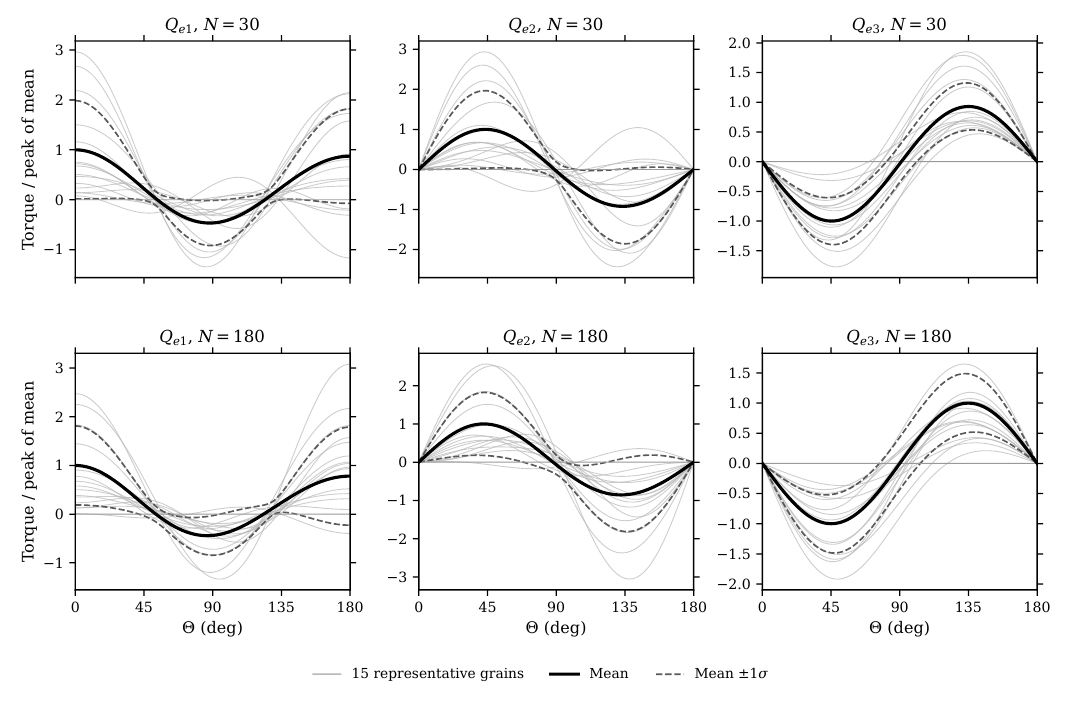}
\caption{Spin-averaged MET torques at $s_d=1$ for the first 30 grains (top) and the full sample (bottom). Gray curves: 15 representative grains. Black: ensemble means. Dashed: $\pm1\sigma$. For the statistical means only, we reverse the signs of $Q_{e1}$ and $Q_{e2}$ when $Q_{e1}(0)<0$. Individual-grain dynamics use the original signs.}
\label{fig:torques180}
\end{figure*}

\begin{figure}[t]
\centering\includegraphics[width=\columnwidth]{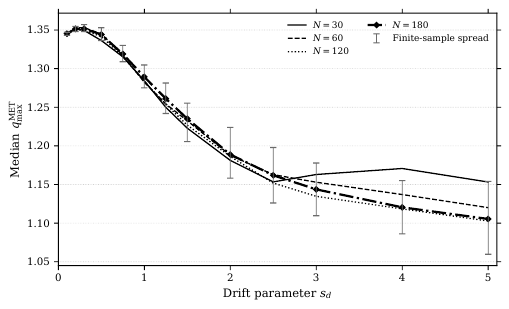}
\caption{Median individual-grain torque ratio $q_{\max}^{\rm MET}$ versus drift speed, for successively larger samples of grain shapes. Error bars show how the estimated median varies when different grain shapes are drawn from the ensemble. They do not represent fluctuations from individual gas impacts. The ratio describes the relative, not absolute, strengths of the two torque components.}
\label{fig:qmax180}
\end{figure}

\begin{figure*}[t]
\centering\includegraphics[width=.98\textwidth]{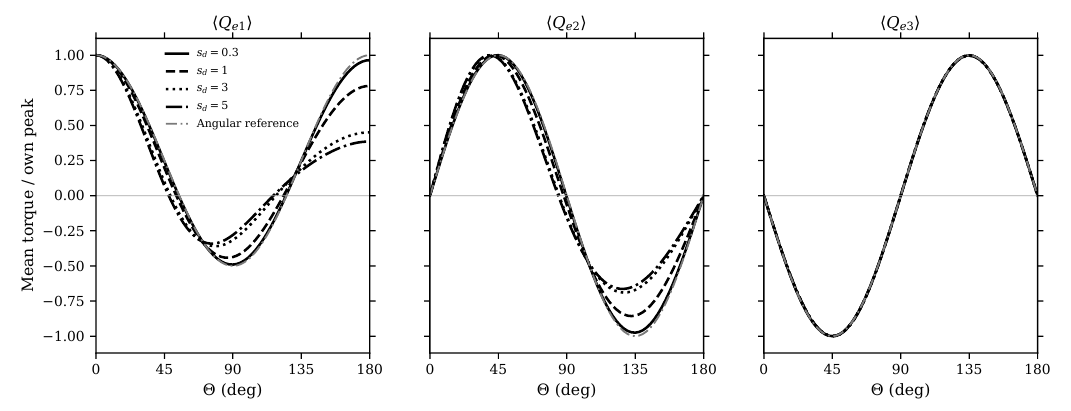}
\caption{Angular profiles of the sample-mean mechanical torque components at four drift speeds. Each profile is normalized by its own peak absolute amplitude. The near-overlap in the $Q_{e3}$ panel shows that the angular profile remains close to $-\sin 2\Theta$. The unnormalized peak magnitude grows from 0.03 at $s_d=0.3$ to 0.3 at $s_d=1$ and 2.6 at $s_d=3$. Gray curves show the leading weak-drift angular forms.}
\label{fig:torquefit180}
\end{figure*}

For a compact comparison with RATs, define $\rho=Q_{e3}^{\max}/\max(Q_{e1}^{\max},Q_{e2}^{\max})$ using the peak magnitude of each torque component for an individual grain. The median MET $\rho$ increases from 1.8 at $s_d=0.3$ to 3.2 at $s_d=1$ and 4.4 at $s_d=3$ (Figure~\ref{fig:met-rat-q3-ratio}). The specified RAT analytical model in Figure~\ref{fig:met-rat-q3-ratio} gives $\rho=0.78$. In the numerical RAT calculations of \citet[Figure~8]{LH2007rat}, some irregular grains have $Q_{e3}^{\max}/Q_{e1}^{\max}>1$. RAT precession and its competition with Larmor precession were also explicitly calculated by \citet[Section~9.1]{LH2007rat}. Large relative precessional torques are therefore possible for both RATs and METs; the numerical MET trend here refers specifically to changing the drift strength.


\subsection{Comparison with AMO mirrors of different inclination}
\label{sec:amo-mirror}

\begin{figure*}[t]
\centering\includegraphics[width=.98\textwidth]{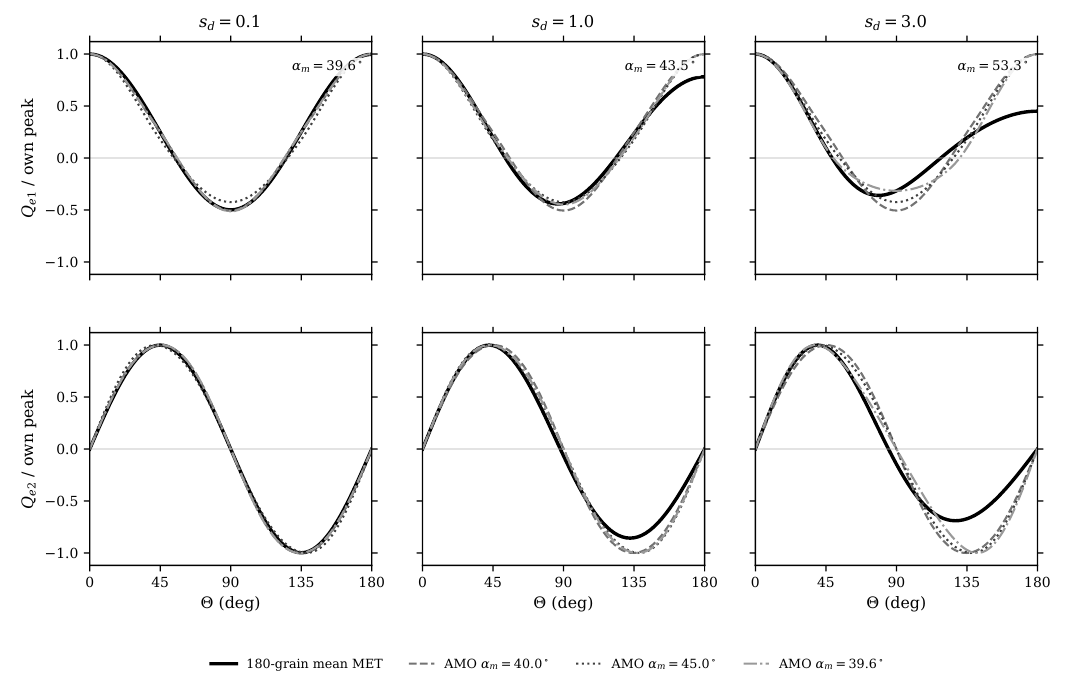}
\caption{Sample-mean MET $Q_{e1}$ and $Q_{e2}$ angular profiles compared with LH07 AMO mirrors. The curves compare the usual $45^\circ$ mirror with an alternative inclination that gives a closer match to the MET angular shape at each illustrated drift speed.  Every curve is normalized by its own peak.  }
\label{fig:amomirror}
\end{figure*}

The RAT analytical model (AMO) represents grain helicity by an inclined mirror attached to an oblate body \citep{LH2007rat}. Its inclination $\alpha_m$ is fixed for a given model, with $45^\circ$ commonly used. Figure~\ref{fig:amomirror} compares spin-averaged AMO and ensemble-mean MET angular profiles, with each component normalized separately. The mirror inclination controls the comparison of angular shapes.

The mean MET profiles resemble RAT AMO more closely than the gas-bombarded single mirror of \citet{LH2007met}. The latter treats one idealized reflecting surface, whereas our irregular grains contain many facets. Averaging the torques after matching their helicity signs gives an AMO-like mean despite the differences among individual grains. Their ensemble mean also changes with $s_d$ (Figure~\ref{fig:torquefit180}).

An AMO mirror near $40^\circ$ best describes the mean angular shapes at weak drift. The comparison inclination rises to about $44^\circ$ at $s_d=1$ and $53^\circ$ at $s_d=3$. Higher drift introduces noticeable departures, and matching separately normalized profiles does not fix their amplitude ratio $q_{\max}$, introduced by \citet{LH2007rat}. 

\section{Precession averaging and stability}\label{sec:stability}

\subsection{Averaging about the dominant external reference axis}

With efficient internal relaxation, write $\mathbf J=j\hat{\mathbf u}$, where $\hat{\mathbf u}$ is the maximum-inertia axis. The rapid-precession reference axis may be the magnetic field $\hat{\mathbf B}$ (Barnett-induced Larmor precession) or the drift direction $\hat{\mathbf k}=\hat{\mathbf v}$ (precession driven by $G$) \citep{DolginovMitrofanov1976, Purcell1979, LH2007met}.

For Barnett magnetization, $\boldsymbol\mu_{\rm Bar}=\chi(0)V\boldsymbol\omega/\gamma_e$, where $\gamma_e$ is the electron gyromagnetic ratio. The Larmor angular frequency and period are \citep[see][]{LH2007rat}
\begin{equation}
 \Omega_B=\frac{\chi(0)VB}{|\gamma_e|I_1},\qquad t_B=\frac{2\pi}{\Omega_B}.\label{eq:barnettperiod}
\end{equation}
For a compact spherical grain with $I_1\propto\rho a^5$ and $V\propto a^3$, $t_B\propto\rho a^2/[\chi(0)B]$. At $a=0.03\,\mu\mathrm m$, $\rho=2.2\,\mathrm{g\,cm^{-3}}$, $\chi(0)=10^{-4}$, and $B=100\,\mu\mathrm G$, the spherical-inertia estimate is $t_B\simeq1$ day. This is much shorter than the illustrative gas-drag time of about 45 yr considered below. Magnetic inclusions can further shorten $t_B$ by increasing the susceptibility \cite{LazarianHoang2008}. 

The mechanical azimuthal torque produces drift-axis precession, as in the mechanical-alignment dynamics of \citet{LH2007met,HoangEtAl2022Disk}. From the azimuthal component $G$, the characteristic frequency is
\begin{equation}
 \Omega_k(\theta)=\frac{\Gamma_0 G(\theta)}{j\sin\theta},\qquad
 t_k(\theta)=\frac{2\pi |j|\sin\theta}{\Gamma_0|G(\theta)|},
\end{equation}
where $j$ is the signed body-axis angular momentum.  Unlike $\Omega_B$, $\Omega_k$ depends on the grain shape, surface-interaction law, drift parameter, instantaneous spin, and orientation. In particular, $\Gamma_0\propto n_gm_gv_{\rm th}^2a^3$, so the mechanical-precession periods differ among the grain shapes. At a suprathermal equilibrium with $|j|\sim\Gamma_0\tau_{\rm drag}|H|$, its characteristic scale is $|\Omega_k|\sim |G/(H\sin\theta)|/\tau_{\rm drag}$, which must be evaluated for each shape. The precession-axis selection is controlled by comparing $|\Omega_B|$ and $|\Omega_k|$.

When the dominant precession period is much shorter than the spin-up and inclination-evolution times, the rapidly changing azimuth can be averaged out. The grain drift and magnetic-field directions are taken as fixed during one precession cycle. We distinguish the two possible reference axes explicitly.

\emph{Fast precession about the magnetic field.} Define
$\psi=\angle(\hat{\mathbf u},\hat{\mathbf B})$,
$\beta=\angle(\hat{\mathbf k},\hat{\mathbf B})$, and let $\phi_B$ be the azimuth of $\hat{\mathbf u}$ about $\hat{\mathbf B}$, measured from the plane containing $\hat{\mathbf k}$ and $\hat{\mathbf B}$. As the grain precesses, $\psi$ remains nearly fixed but its angle to the drift changes according to
\begin{equation}
 \cos\Theta=\cos\psi\cos\beta
       +\sin\psi\sin\beta\cos\phi_B.
 \label{eq:met-angle-B}
\end{equation}
The mechanical torque at each azimuth is
$\mathbf Q=H(\Theta)\hat{\mathbf u}
 +F(\Theta)\hat{\mathbf e}_{\Theta}
 +G(\Theta)\hat{\mathbf e}_{\phi}$, where
$\hat{\mathbf e}_{\Theta}=(\cos\Theta\hat{\mathbf u}-\hat{\mathbf k})/\sin\Theta$
and $\hat{\mathbf e}_{\phi}=\hat{\mathbf k}\times\hat{\mathbf u}/\sin\Theta$.
The polar direction relative to the magnetic field is
$\hat{\mathbf e}_{\psi}=(\cos\psi\hat{\mathbf u}-\hat{\mathbf B})/\sin\psi$.
Projection onto the spin and magnetic-polar directions followed by averaging gives
\begin{align}
 H_B(\psi;\beta)&=\frac{1}{2\pi}\int_0^{2\pi}
                     H(\Theta)\,d\phi_B,\label{eq:HB-explicit}\\
 F_B(\psi;\beta)&=\frac{1}{2\pi}\int_0^{2\pi}
 F(\Theta)\,\frac{\cos\beta-\cos\psi\cos\Theta}
 {\sin\psi\sin\Theta}\,d\phi_B.
 \label{eq:FB-explicit}
\end{align}
The factor multiplying $F$ is $\hat{\mathbf e}_{\Theta}\cdot\hat{\mathbf e}_{\psi}$. The $G$ projection is odd in $\phi_B$ and averages to zero. $G$ drives precession but not secular magnetic-polar alignment in this limit. We take continuous axial limits at $\psi=0,\pi$. For $j<0$, the angle between $\mathbf J$ and $\mathbf B$ is $\pi-\psi$.

\emph{Fast precession about the drift direction.} When mechanical precession dominates, the external reference axis is $\hat{\mathbf A}=\hat{\mathbf k}$. During rotation about $\hat{\mathbf k}$ the angle $\Theta$ to the gas drift is \emph{constant}. Unlike the magnetic-axis case, the grain does not pass through different drift-relative angles. The secular functions are simply
\begin{equation}
 H_k(\Theta)=H(\Theta),\qquad F_k(\Theta)=F(\Theta).
 \label{eq:met-k-average}
\end{equation}
Here $G$ supplies the azimuthal precession, while $F$ changes the grain--drift inclination. For oblique $\mathbf B$, magnetic dissipation must be averaged separately about $\hat{\mathbf k}$. For the magnetic-damping law defined in Section~\ref{sec:delta-environments}, it adds a polar drift $\dot\Theta_{\rm mag}=-\delta_m P_2(\cos\beta)\sin\Theta\cos\Theta/\tau_{\rm drag}$, rather than the fast-$B$ term $-\delta_m\sin\psi\cos\psi/\tau_{\rm drag}$. The numerical $R_B$ results below are for fast $B$ precession and use Eqs.~\eqref{eq:HB-explicit}--\eqref{eq:FB-explicit}. Both averages require precession to be faster than secular spin and inclination evolution.

For signed body-axis angular momentum $j=I_1\omega$, the high-spin equations have the same spin-up, drag, and aligning-torque structure as the RAT equations of \citet{LH2007rat,HoangLazarian2016}, with the calculated MET functions replacing the RAT functions:
\begin{align}
 \dot j&=\Gamma_0H_A(\psi)-j/\tau_{\rm drag},\\
 \dot\psi&=\Gamma_0F_A(\psi)/j.
\end{align}
At equilibrium, $F_A(\psi_*)=0$ and $j_*=\Gamma_0\tau_{\rm drag}H_A(\psi_*)$. We write $|J_*|=|j_*|$. Spin perturbations decay on the damping time, while the angular growth rate is $\lambda_{\psi}=F_A'(\psi_*)/[\tau_{\rm drag}H_A(\psi_*)]$.

Stability requires perturbations in both spin and inclination to decay.

We retain either sign of $j$, test axial endpoints with one-sided derivatives, and identify a stable high-$J$ attractor when $|j_*|/J_{\rm th}\ge3$. The adopted nominal $Q_{\rm drag}=5$, reference $a=0.03\,\mu$m, and $\rho=2.2\,\mathrm{g\,cm^{-3}}$ provide an illustrative spin normalization.

\paragraph{Axial and tilted equilibria.}
An axial equilibrium has the grain angular momentum parallel or antiparallel to the reference axis about which it precesses. A \emph{tilted equilibrium} instead has a stationary, nonzero inclination to that axis in the precession-averaged dynamics: $0<\psi_*<\pi$ for rapid precession about $\mathbf B$, or $0<\Theta_*<\pi$ for rapid precession about the drift direction. The angular momentum still precesses around that axis. Only its inclination and magnitude are stationary. Without magnetic dissipation, a tilted state is possible wherever the averaged polar MET torque vanishes, $F_A(\psi_*)=0$, and the spin-up balances gas drag. With magnetic dissipation, the polar MET torque can instead balance the magnetic aligning torque at a nonzero inclination (Section~\ref{sec:magneticMET}). A root of the stationary equations is an attractor only if both the spin and angular perturbations decay. Because $P_2(\cos\psi_*)$ or $P_2(\cos\Theta_*)$ need not equal unity, tilted states also contribute differently from axial states to the ensemble Rayleigh factor.

\subsection{Torque ratio, drift geometry, and high-$J$ attractors}
\label{sec:qmax-highJ}
The significance of $q_{\max}$ follows the RAT analysis of \citet{LH2007rat}, namely, the $Q_{e1}$ component favors axial spin-up, whereas $Q_{e2}$ can drive the grain away from that orientation. Their competition depends on the angle between the directed forcing and the magnetic field. In the RAT case this is the radiation--field angle $\psi_{\rm RAT}$. In the MET case it is the drift--field angle $\beta$. We use $q_{\max}^{\rm MET}$ defined by Eq.~\eqref{eq:met-qmax}. It is calculated for each grain, not from the ratio of two separately averaged curves.

For clarity, adopt the lowest-order AMO-like angular template motivated by \citet{LH2007rat}:
\begin{equation}
 Q_{e1}=q\,\frac{5\cos^2\Theta-2}{3},\qquad
 Q_{e2}=\sin2\Theta,\qquad q=q_{\max},
 \label{eq:qmax-amo-template}
\end{equation}
with the overall torque scale suppressed. If $\hat{\mathbf k}\parallel\hat{\mathbf B}$, the axial state has $H(0)=q$ and $F'(0)=2-q$. In the absence of magnetic relaxation it is stable against a small inclination perturbation for $q>2$. With the magnetic relaxation parameter $\delta_m$ defined in \S~\ref{sec:delta-environments}, the axial condition becomes (see Eq. (\ref{eq:qmax-amo-template}):
\begin{equation}
 \frac{F_B'(0)}{H_B(0)}<\delta_m,
 \qquad q>\frac{2}{1+\delta_m}
 \quad (\beta=0).
 \label{eq:qmax-axis-criterion}
\end{equation}
Thus magnetic dissipation can preserve a rapidly rotating state which would otherwise be unstable to a small change in orientation. It does so by damping the departure from the magnetic axis.

For oblique drift, Eqs.~\eqref{eq:HB-explicit}--\eqref{eq:FB-explicit} give $H_B(0)=H(\beta)$ and
\begin{equation}
 F_B'(0)=\tfrac12\bigl[F'(\beta)+F(\beta)\cot\beta\bigr].
 \label{eq:qmax-beta-criterion}
\end{equation}
For the AMO-like family, the stability boundary becomes
\begin{equation}
 \delta_{m,{\rm crit}}^{\rm AMO}(q,\beta)=
 \frac{(2-q)+(10q/3-4)\sin^2\beta}
 {q+(2-5q/3)\sin^2\beta},
 \label{eq:qmax-amo-deltacrit}
\end{equation}
provided $H(\beta)\ne0$. At $\beta=0$, the axial boundary moves from $q>2$ without magnetic relaxation to $q>2/3$ at $\delta_m=2$. As in the RAT analysis of \citet{LH2007rat}, magnetic damping widens the angular-stability domain. The measured $q_{\max}\simeq1.1$--1.4 therefore makes angular stability sensitive to drift geometry and magnetic relaxation. 

\subsection{An analytical weak-drift check}
The weak-drift limit provides an analytical constraint on axial stability. We obtain the following expansion directly from our collision model, in analogy with the linear-response treatment of mechanical torques in \citet{DasWeingartner2016}:
\begin{align}
 \Gamma_i(\hat{\mathbf v},s_d)
   &=\Gamma_i^{(0)}+\Gamma_0s_dD_{ij}\hat v_j+O(s_d^2),\\
 D_{ij}&=\left.\frac{1}{\Gamma_0}
      \frac{\partial\Gamma_i}{\partial(s_d\hat v_j)}\right|_{s_d=0}.
\end{align}
Here $\mathbf D$ is the dimensionless \emph{linear mechanical-torque response tensor}. It measures the torque produced by a directed grain--gas velocity. Its zeroth-order torque vanishes for our grain shapes, while diagonal response is sensitive to the handedness of the grain--surface interaction.

After spin averaging, axial symmetry allows a longitudinal response
$D_\parallel=\hat{\mathbf u}\cdot\mathbf D\hat{\mathbf u}$ and two equal
transverse diagonal responses $D_\perp$. An antisymmetric transverse response
may also be present, but it contributes to azimuthal torque rather than to the
polar torque considered here. Its contraction over the torque and drift indices vanishes by antisymmetry, even before spin averaging. Thus $\operatorname{tr}\mathbf D=0$, so
\begin{equation}
 D_\parallel+2D_\perp=0,
 \qquad D_\perp=-\frac12D_\parallel.
\end{equation}
The factor of two reflects the two directions perpendicular to the grain axis.
A nonzero $D_\parallel$ describes a torque along the grain axis generated by a
weak directed gas flow. 

Projecting the spin-averaged response tensor onto the scattering frame of Eq.~(\ref{eq:met-scattering-basis}) gives, to first order in drift speed,
\begin{align}
 Q_{e1}(\Theta)&=s_dD_\parallel\,P_2(\cos\Theta)+O(s_d^2),\\
 Q_{e2}(\Theta)&=\frac{3}{4}s_dD_\parallel\sin2\Theta+O(s_d^2).
 \label{eq:met-Qe-weak}
\end{align}
These are the weak-drift angular shapes used for the AMO comparison. Applying Eq.~(\ref{eq:met-Qe-projection}) then yields
\begin{align}
 H(\Theta)&=s_dD_\parallel\cos\Theta+O(s_d^2),\\
 F(\Theta)&=\frac12s_dD_\parallel\sin\Theta+O(s_d^2).
\end{align}
Thus $F'(0)=s_dD_\parallel/2+O(s_d^2)$ and
$H(0)=s_dD_\parallel+O(s_d^2)$. Provided $D_\parallel\ne0$, their ratio is
\begin{equation}
 \left.\frac{F'(\Theta)}{H(\Theta)}\right|_{\Theta=0}
 =+\frac12+O(s_d).
 \label{eq:weak}
\end{equation}
The sign is independent of the sign of $D_\parallel$. Reversing the handedness grains
reverses both $F'$ and $H$.

Recall that
$\dot\Theta=\Gamma_0F(\Theta)/j$ and
$j_*=\Gamma_0\tau_{\rm drag}H(0)$ at an axial stationary point.
Linearizing about $\Theta=0$ gives
\begin{equation}
 \delta\dot\Theta
 =\frac{1}{\tau_{\rm drag}}\frac{F'(0)}{H(0)}\,
 \delta\Theta
 =\left[\frac{1}{2\tau_{\rm drag}}+O(s_d/\tau_{\rm drag})\right]\delta\Theta.
\end{equation}
The angular perturbation therefore grows. The $+1/2$ ratio follows from the weak-drift collision response.

\section{Numerical Rayleigh reduction factors}
\label{sec:results}
\subsection{High-$J$ population model}
We calculate alignment for the grain sample at $s_d=0.3$, 1, and 3, using 24 grain-spin phases and 73 angles $0\leq\Theta\leq\pi$. The first 30 grains provide a direct check against the earlier calculation. The pure-collision limits use the same surface integration. Rapid magnetic-precession averages use 128 azimuthal samples and 361 angular samples. We evaluate $\beta=0^\circ$--$90^\circ$, including a dense grid near transverse drift, and $\delta_m=0$, $0.25$, $0.5$, 1, $1.5$, 2, 5, and 10. To resolve the weak-relaxation regime we additionally sample $0\leq\delta_m\leq0.5$ in increments of $0.025$ at eleven drift--field angles and all three drift strengths. The grain realizations are defined by the seed and shape-generation procedure in Section~2.

Only stable fixed points with $|J_*|/J_{\rm th}\geq3$ contribute to the alignment factor. For each grain with stable high-$J$ attractors, we use the one with the largest $|J_*|$ and assume that it is occupied in the long-time limit.  Grains without stable high-$J$ attractors have zero alignment by definition: low-$J$ alignment is outside the calculation. For equal grain-shape weights and perfect internal alignment, the ensemble Rayleigh factor is
\begin{align}
 R_{B,\infty} &= \frac1N\sum_{i=1}^{N}r_i, \label{eq:rayleigh}\\
 r_i &= \begin{cases}
 P_2(\cos\psi_{*,i}), & \text{stable high-}J,\\
 0, & \text{otherwise},
 \end{cases}\\
 P_2(x)&=\frac12(3x^2-1),\qquad |J_*|\geq3J_{\rm th}.
\end{align}
We write $R_B\equiv R_{B,\infty}$ for the saturated high-$J$ ensemble measure in the selected-state approximation. To relate it to alignment attained during finite exposure, let $P_{H,i}(t)$ be the probability of occupation of the selected high-$J$ state of grain $i$. Then the high-$J$ contribution is
\begin{equation}
 R_B^{(H)}(t)=\frac{1}{N}\sum_{i\in\mathcal H_B}P_{H,i}(t)P_2(\cos\psi_{*,i}),
 \label{eq:met-finite-R}
\end{equation}
where $\mathcal H_B$ denotes grains possessing the selected stable high-$J$ state. This expression assumes perfect internal alignment in that state and neglects other states. The tabulated factors correspond to setting $P_{H,i}=1$ for these grains. The factor $P_{H,i}$ refers to actual occupation after any unsuccessful low-$J$ excursions and recapture, rather than to the probability of a single boundary crossing. Persistent drift can permit the saturated limit, but the finite-time result depends on the distribution of successful first-arrival times. Near-thermal rotation can persist for much longer than gas damping.

\subsection{Drift-axis high-$J$ occupation}\label{sec:drift-highJ}
When the drift direction itself is the reference axis, the spin and angular torques are $H_k(\Theta)=H(\Theta)$ and $F_k(\Theta)=F(\Theta)$.  For each grain we define
\begin{equation}
 f_{H,k}=\frac{N_{H,k}}{N},\qquad
 R_{k,\infty}=\frac1N\sum_{i\in {\cal H}_k}P_2(\cos\Theta_{*,i}),
 \label{eq:drift-occupation}
\end{equation}
where ${\cal H}_k$ contains grains with a stable high-$J$ state satisfying $|J_*|\geq3J_{\rm th}$ with $N$ defined in Section~2. In the selected-state approximation, if every qualifying attractor is axial ($\Theta_*=0$ or $\pi$), then $R_{k,\infty}=f_{H,k}$. Tilted high-$J$ states make the two quantities different. Neither quantity is a finite-time probability of capture from low $J$.

For $\beta=0$, the magnetic field and drift direction coincide. In this special geometry the precession-averaged magnetic-axis functions reduce \emph{exactly} to $H_k$ and $F_k$, whether fast precession is supplied by the magnetic field or by the mechanical torque. The derivation, including the finite-$\beta$ torque averages, is given in Appendix~\ref{app:B-average}. Consequently the original magnetic-axis result at $\beta=0$ and $\delta_m=0$ is already a drift-axis Rayleigh factor for the same grain ensemble. In particular, the original calculation at $s_d=1$ gives $R_{k,\infty}=R_B(\beta=0,\delta_m=0)\simeq0.07$. This Rayleigh factor alone does not determine the number of high-$J$ grains if any stable equilibria are tilted.

To estimate $N_{H,k}$ independently, we generated another sample of the same size with the stated spherical-harmonic prescription (90 each with $\alpha=2$ and 3), the 50/50 collision rule, and spin-averaged surface torques.  We used the $a=0.03\,\mu$m, $\rho=2.2\,\mathrm{g\,cm^{-3}}$, $T_g=10$ K reference and the gas-damping normalization of Section~2. We counted stable positive- and negative-spin equilibria separately before selecting the one with largest $|J_*|$ in each grain. The resulting fractions and drift-axis Rayleigh factors appear in Table~\ref{tab:drift-highJ-recon}.

\begin{deluxetable}{rrrr}
\tablecaption{Drift-axis high-$J$ population in an independent grain sample\label{tab:drift-highJ-recon}}
\tablehead{\colhead{$s_d$}&\colhead{$N_{H,k}/N$}&\colhead{$f_{H,k}$}&\colhead{$R_{k,\infty}$}}
\startdata
0.3 & 7/180  & 0.04 & 0.04 \\
1 & 16/180 & 0.09 & 0.09 \\
3 & 22/180 & 0.12 & 0.08 \\
\enddata
\tablecomments{All numerical entries use the independently reconstructed ensemble. The fraction $f_{H,k}$ counts grains with a selected stable high-$J$ state, whereas $R_{k,\infty}$ also weights each state by $P_2(\cos\Theta_*)$. At $s_d=3$, some stable equilibria are tilted relative to the drift, giving $R_{k,\infty}\ne f_{H,k}$.}
\end{deluxetable}

At $s_d=3$, the ratio $R_{k,\infty}/f_{H,k}\simeq0.69$ is the mean $P_2(\cos\Theta_*)$ among the selected high-$J$ states, rather than a second estimate of their number. It reflects their nonaxial equilibrium inclinations; individual tilted angles need not be the same.

The independently reconstructed median $q_{\max}$ agrees with the original ensemble to about 0.01 across the three drift speeds. Doubling the grain-spin and inclination sampling leaves the three reconstructed occupation counts unchanged. The close agreement in torque-ratio statistics supports the reconstruction as a useful independent numerical test, but does not replace the original per-grain arrays for determining their exact high-$J$ counts.

\subsection{Angular distributions and ensemble values}
\begin{figure*}[t]
\centering\includegraphics[width=.98\textwidth]{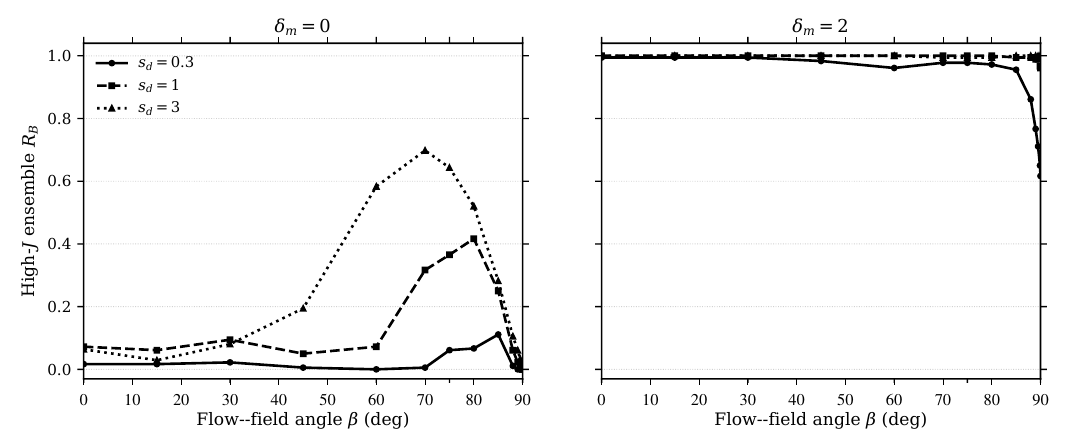}
\caption{Saturated magnetic-axis Rayleigh factor of the default model against flow--field angle $\beta$ for $s_d=0.3$, 1, and 3 (line styles), without extra magnetic relaxation (left) and with $\delta_m=2$ (right). }
\label{fig:scan}
\end{figure*}

The enhancement of high-$J$ occupation with magnetic relaxation is illustrated by the perpendicular-drift limit. At $s_d=1$, $R_{B,\infty}$ rises from zero without enhanced magnetic relaxation to about 0.96 at $\delta_m=2$. For oblique drift near $75^\circ$, appreciable alignment occurs even at $\delta_m=0$. The increase primarily reflects more grains acquiring stable rapidly rotating states. At exactly perpendicular drift and $\delta_m=2$, alignment also becomes stronger as the drift parameter rises. The grain-size dependence enters through the drift speed and magnetic relaxation, evaluated below for molecular-cloud and disk conditions.

Most changes in $R_B$ arise because more grains acquire stable, nearly field-aligned high-$J$ attractors.

\section{MET alignment with weak and enhanced magnetic relaxation}\label{sec:magneticMET}
\subsection{Magnetic-relaxation parameter and high-$J$ stability}
We distinguish weak ($0\leq\delta_m\leq0.5$), transitional ($0.5<\delta_m\leq1$), and enhanced ($\delta_m>1$) magnetic relaxation. These ranges organize the calculated high-$J$ stability results. With reduced signed spin $\tilde j=j/(\Gamma_0\tau_{\rm drag})$ and $t'=t/\tau_{\rm drag}$,
\begin{align}
 \frac{d\tilde j}{dt'}&=H_B(\psi)-\tilde j[1+\delta_m\sin^2\psi],\\
 \frac{d\psi}{dt'}&=F_B(\psi)/\tilde j-\delta_m\sin\psi\cos\psi,
 \label{eq:magflow}
\end{align}
For an axial state at $\psi_*=0$, the angular stability condition is $F_B'(0)/H_B(0)<\delta_m$ when $H_B(0)\ne0$. At a tilted equilibrium, $0<\psi_*<\pi$, the stationary spin and inclination satisfy
\begin{align}
 \tilde j_* &= \frac{H_B(\psi_*)}{1+\delta_m\sin^2\psi_*},\label{eq:met-tilted-spin}\\
 F_B(\psi_*) &= \delta_m\tilde j_*\sin\psi_*\cos\psi_*.
 \label{eq:met-tilted-angle}
\end{align}
The second equation balances mechanical and magnetic aligning torques. At $\delta_m=0$, it gives $F_B(\psi_*)=0$. We classify a tilted stationary point by perturbing both $\tilde j$ and $\psi$ and evaluating the eigenvalues of the resulting two-dimensional Jacobian. Both eigenvalues must have negative real parts for a stable attractor. A tilted state can therefore be stable or unstable, and its Rayleigh contribution is $P_2(\cos\psi_*)$, not the unit contribution of an axial state. Only stable high-$J$ solutions enter $R_B$.

\subsection{Weak magnetic relaxation: $0\leq\delta_m\leq0.5$}\label{sec:low-delta}

\begin{figure*}[t]
\centering\includegraphics[width=.98\textwidth]{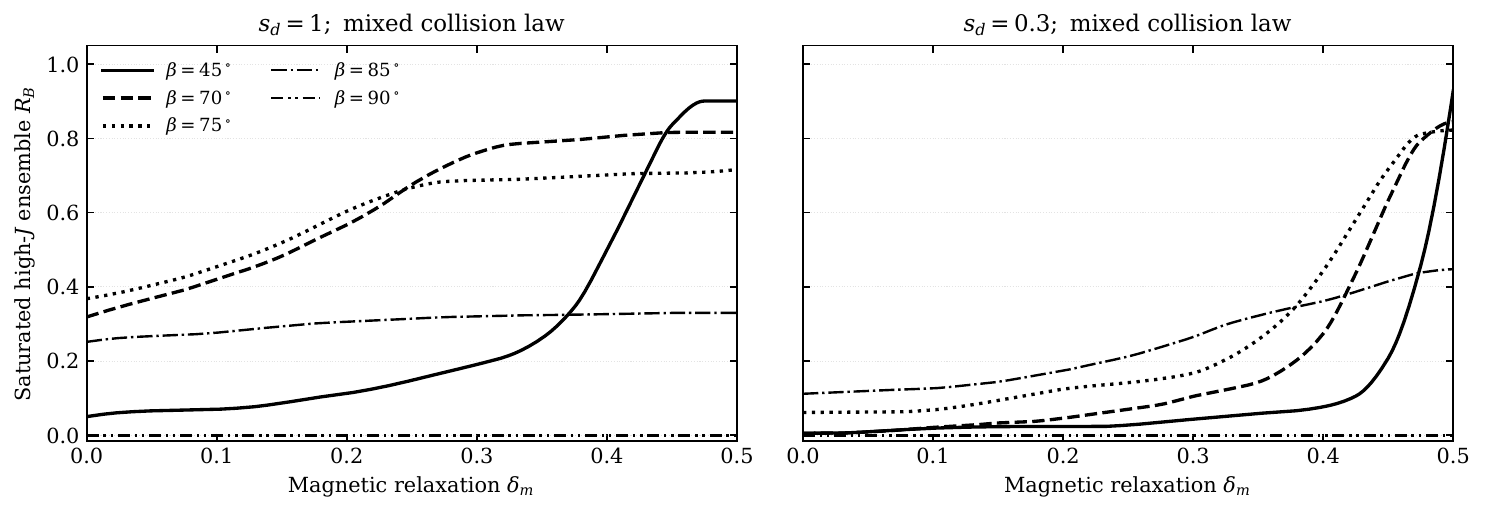}
\caption{Saturated high-$J$ Rayleigh factor $R_B$ against magnetic relaxation for transonic ($s_d=1$, left) and subsonic ($s_d=0.3$, right) drift at different flow--field angles $\beta$. The curves have been smoothed from the sample results to display the systematic trend. No stable high-$J$ state contributes zero. The smoothing is for presentation and does not represent additional numerical precision.}
\label{fig:lowdelta}
\end{figure*}

We calculate the mixed-law high-$J$ states for weak magnetic relaxation, including the onset of stable high-$J$ alignment (Figure~\ref{fig:lowdelta}).

For oblique drift, even moderate magnetic relaxation can increase the saturated $R_B$ substantially. The strength and onset of alignment depend on the flow--field angle. Near exactly perpendicular drift, appreciable alignment generally requires stronger magnetic relaxation than shown in Figure~\ref{fig:lowdelta}. This is a change in the stability of the high-$J$ equilibria, not merely a rotation of an already aligned population. The local angular growth or decay rate near an axial equilibrium follows from Eq.~\eqref{eq:magflow}:
\begin{equation}
 \lambda_{\perp}\tau_{\rm drag} = \frac{F_B'(0)}{H_B(0)}-\delta_m,
 \qquad \delta_{m,\mathrm{crit}}=\frac{F_B'(0)}{H_B(0)}.
 \label{eq:lowdelta-stability}
\end{equation}
We also test the opposite axial orientation and the coupled spin--inclination stability of tilted equilibria. The nontransverse weak-drift limit has a characteristic onset near $\delta_m\simeq0.5$, whereas perpendicular drift requires a stronger magnetic response because its leading-order torque geometry differs. The corresponding eigenvalue determines the local restoring time.

The increase in $R_B$ mainly reflects the appearance of stable high-$J$ attractors in more grain shapes. Gas impacts continually perturb the direction of $\mathbf J$ about its equilibrium inclination. A small perturbation in the stability analysis means an infinitesimal angular displacement $\Delta\psi$ from a stationary state $\psi_*$, not a prescribed collision of a particular strength. Near a stable axial state at $\psi_*=0$, the unforced displacement decays as $\Delta\psi(t)\propto\exp(\lambda_\perp t)$, with the local restoring time
\begin{equation}
 \tau_{\rm restore} = \frac{1}{|\lambda_\perp|}
 =\frac{\tau_{\rm drag}}{\delta_m-F_B'(0)/H_B(0)}
 \quad (\lambda_\perp<0).
 \label{eq:met-local-restoring-time}
\end{equation}
For a tilted state the spin and inclination perturbations are coupled, and the slowest decaying eigenmode of the two-dimensional Jacobian determines the local restoring time. Close to a stability threshold this time becomes long because the restoring eigenvalue approaches zero. It measures relaxation \emph{after} a small disturbance of an already occupied high-$J$ state. It does not measure the stochastic waiting time required for a grain trapped at low $J$ to cross the attraction-basin boundary and reach high $J$. The reported $R_{B,\infty}$ values remain saturated reference limits.

\subsection{Transition region: $0.5<\delta_m\leq1$}
For example, for exactly transverse drift at $s_d=1$, the ensemble gives $R_B=0$ below $\delta_m=0.5$ and only $R_B\simeq0.09$ at $\delta_m=1$, while oblique cases acquire stable high-$J$ states at smaller $\delta_m$.

\subsection{Enhanced magnetic relaxation: $\delta_m>1$}
\begin{figure}[t]
\centering
\includegraphics[width=\columnwidth]{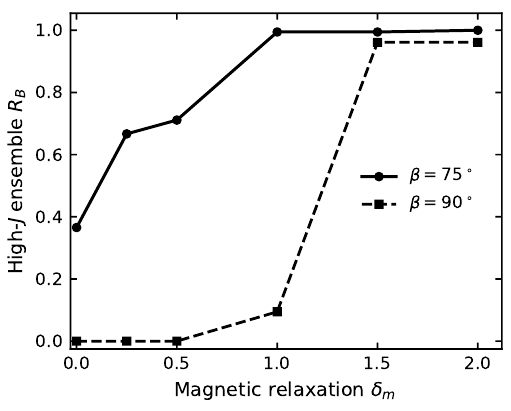}
\caption{High-$J$ ensemble $R_B$ at $s_d=1$, for $\beta=75^\circ$ (solid circles) and $90^\circ$ (dashed squares). The curves connect the computed $\delta_m$ grid points. At $\delta_m=5$ and 10 they remain at their respective plateaus, $R_B\simeq1$ and 0.96.}
\label{fig:rbdelta}
\end{figure}

Following the magnetic-dissipation extension to RAT alignment in \citet{LazarianHoang2008}, this is the enhanced-relaxation counterpart for METs. The mixed model shows a rapid transition toward axial high-$J$ alignment at moderate magnetic damping, although the threshold depends on the drift direction and strength. For perpendicular drift at $s_d=1$, we calculate $R_B=0$, 0.09, and 0.96 for $\delta_m=0$, 1, and 2. At $\delta_m=2$ the limiting sticking and specular prescriptions give $R_B=0.87$ and 0.97, respectively. Enhanced magnetic relaxation therefore produces strong high-$J$ alignment even for the ideal transverse ambipolar-drift limit.

\subsection{Representative rotational trajectories}\label{sec:portraits}
The phase portraits show stable equilibria and saddles for two grains. Axial points at $\psi=0^\circ$ and $180^\circ$ are plotted at their actual positions.

\begin{figure*}[t]
\centering\includegraphics[width=.97\textwidth]{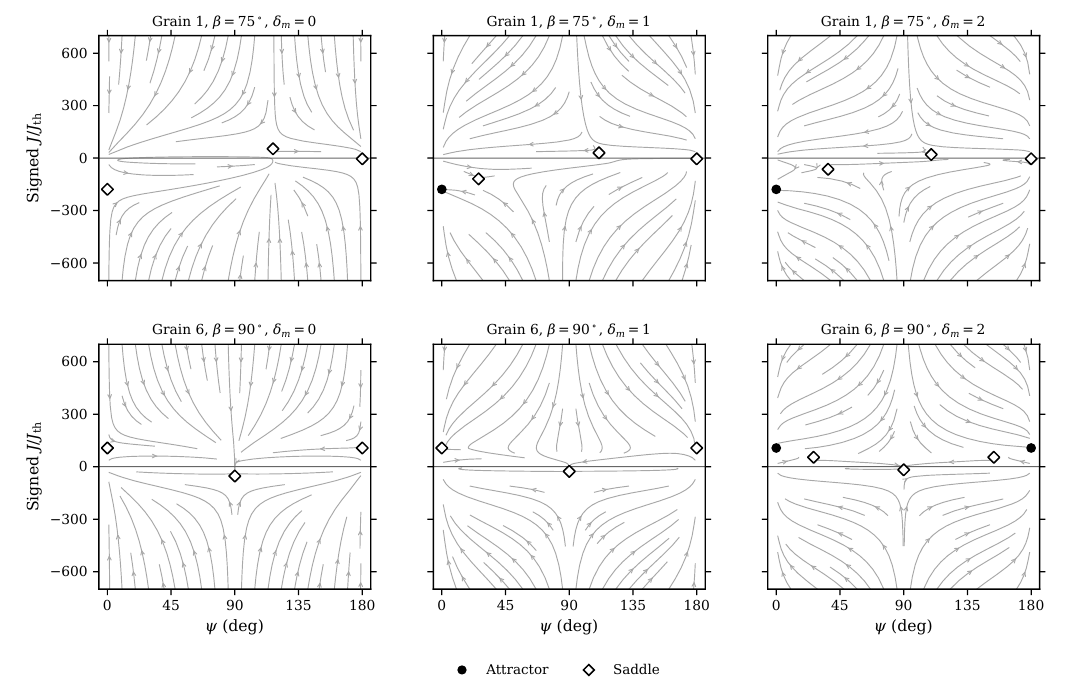}
\caption{Rotational trajectories of grains 1 and 6 at $s_d=1$. Columns show $\delta_m=0$, 1, and 2. Rows correspond to $\beta=75^\circ$ and $90^\circ$. Thin gray curves show how grain spin and orientation change. Filled circles mark \emph{attractors}, toward which nearby trajectories converge. Open diamonds mark \emph{saddles}, which attract along some directions but repel along others. Axial states at $\psi=0^\circ$ and $180^\circ$ are shown at their actual positions. States near zero spin do not model thermal rotation or stochastic changes between low- and high-$J$ states.}
\label{fig:portraits}
\end{figure*}

\begin{figure*}[t]
\centering\includegraphics[width=.98\textwidth]{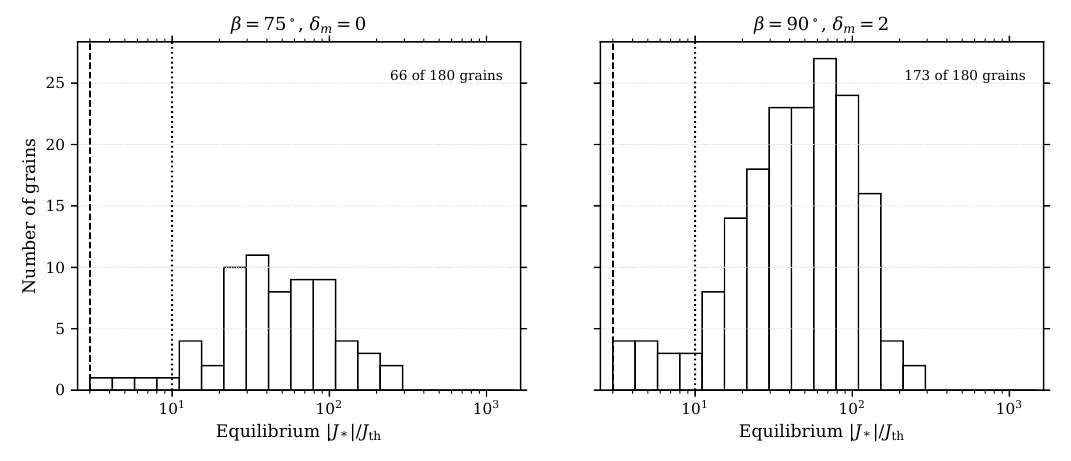}
\caption{Distribution of the equilibrium suprathermal spin $|J_*|/J_{\rm th}$ for high-$J$ states contributing to the mixed-law Rayleigh factor, at $s_d=1$. Left: oblique drift without magnetic relaxation, $\beta=75^\circ$. Right: exactly perpendicular drift with enhanced relaxation $\delta_m=2$. The vertical guides distinguish the adopted threshold $3J_{\rm th}$ and stronger $10J_{\rm th}$ rotation.}
\label{fig:Jrobust}
\end{figure*}

A large $R_B$ is more robust when the selected high-$J$ equilibria also have $|J_*|\gg J_{\rm th}$: a typical random collision then produces a smaller fractional angular-momentum perturbation. The distribution in Figure~\ref{fig:Jrobust} therefore accompanies the alignment factor. For sustained drift, we adopt the high-$J$ occupation prescription defined in Section~2.

\subsection{Near-exact transverse drift}\label{sec:neartransverse}

The transition near perpendicular drift is sharp. At $s_d=1$ and $\delta_m=1$, the mixed law gives $R_B\simeq0.88$ at $\beta=89^\circ$, but about 0.09 at $90^\circ$. At $\delta_m=2$, it gives $R_B\simeq0.96$ even at $90^\circ$: strong magnetic relaxation removes the need for a field-parallel drift component.

\section{Comparison of RATs and METs}\label{sec:RATcompare}
\subsection{Torque geometry and high-$J$ capture}
RAT and MET equations have the same spin-up and orienting-torque structure once the respective torque components are supplied. This resemblance does not make their individual attraction basins identical: MET torques change with grain shape, drift strength, and gas--surface physics, whereas the incident radiation defines the RAT geometry. The high-$J$ MET states in this paper are therefore calculated from each grain's torque functions. The finite-time occupation prescription is given in Eq.~\eqref{eq:met-finite-R}.

\subsection{Relative spin-up torques and grain-size limits}\label{sec:cloud-met-rat}
For gas-damped grains, the spin-up torque in the rapid-magnetic-precession model is $\Gamma_{\rm MET,spin}=n_gm_gv_{\rm th}^2a^3|H_B|$, where $H_B$ is the calculated spin-up coefficient for the chosen grain and drift geometry. The radiative spin-up torque can be expressed approximately as $\Gamma_{\rm RAT,spin}\simeq\gamma_{\rm rad}u_{\rm rad}a^2\bar\lambda\,\bar Q_\Gamma/2$, where $\gamma_{\rm rad}$ is the radiation-field anisotropy, $u_{\rm rad}$ its energy density, $\bar\lambda$ the characteristic wavelength, and $\bar Q_\Gamma$ the RAT spin-up efficiency \citep{LH2007rat,HoangLazarian2014precession}. Their ratio is
\begin{equation}
 \frac{\Gamma_{\rm MET,spin}}{\Gamma_{\rm RAT,spin}}
 \simeq\frac{4n_g k_{\rm B}T_g a}{\gamma_{\rm rad}u_{\rm rad}\bar\lambda}
 \frac{|H_B(s_d,\beta)|}{\bar Q_\Gamma(a)}.
 \label{eq:cloud-met-rat-ratio}
\end{equation}
For $\bar\lambda>1.8a$, the long-wavelength RAT fit of \citet{LH2007rat} gives $\bar Q_\Gamma\simeq0.4(1.8a/\bar\lambda)^3$. Equation~\eqref{eq:cloud-met-rat-ratio} then gives $\Gamma_{\rm MET,spin}/\Gamma_{\rm RAT,spin}\propto |H_B|\bar\lambda^2/a^2$ at fixed gas conditions and radiation field. For $a\gg\bar\lambda$, $\bar Q_\Gamma\simeq0.4$ and the ratio instead grows as $|H_B|a/\bar\lambda$: RAT spin-up becomes less effective relative to MET spin-up for large grains. Attenuation of the radiation field further favors METs by reducing $u_{\rm rad}$. Whether this spin-up produces alignment is determined by the stable high-$J$ attractors calculated in Sections~\ref{sec:stability} and \ref{sec:magneticMET}.

\subsection{MET scaling with grain size and drift}\label{sec:met-size-scaling}
For geometrically similar grains in the free-molecular collision regime, the normalization $\Gamma_0=n_gm_gv_{\rm th}^2a^3$ from Section~\ref{sec:geometry} gives $\Gamma_{\rm MET}\propto a^3$ at fixed drift parameter and gas properties. The dimensionless coefficient depends on shape, orientation, and gas--surface scattering.

For weak drift, the torque calculation of Section~\ref{sec:stability} gives a linear term in $s_d$. At a specified orientation in the magnetic-axis model, we write the spin-up component as $|H_B|\simeq C_H(\beta)s_d$, where $C_H$ is the absolute value of the corresponding weak-drift coefficient for a given grain. Provided that this coefficient does not vanish by symmetry or cancellation,
\begin{equation}
 \Gamma_{\rm MET,spin}\simeq n_gm_gv_{\rm th}a^3
 C_H\,v_{\rm drift}\quad (s_d\ll1).
 \label{eq:met-weak-size}
\end{equation}
The torque is therefore proportional to the grain volume and to the relative drift speed when the linear coefficient is nonzero. An especially important exception occurs in the exactly transverse ambipolar-drift geometry. The weak-drift result $H(\theta)=s_dD_\parallel\cos\theta+O(s_d^2)$, combined with precession averaging and the relation $\langle\cos\theta\rangle=\cos\psi\cos\beta$, implies
\begin{equation}
 H_B(\psi,\beta)=s_dD_\parallel\cos\psi\cos\beta+O(s_d^2).
 \label{eq:met-weak-transverse}
\end{equation}
Thus, at $\beta=90^\circ$, the first-order magnetic-axis spin-up torque vanishes: the leading nonzero contribution can be quadratic in the drift speed. If its coefficient $C_{H,2}\equiv\lim_{s_d\to0}|H_B|/s_d^2$ is nonzero, then $\Gamma_{\rm MET,spin}\propto n_gm_ga^3v_{\rm drift}^2$ and $|J_*|/J_{\rm th}\propto a^{3/2}C_{H,2}s_d^2$ in this special weak-drift geometry. The strong alignment found numerically at $s_d=1$ and $\beta=90^\circ$ therefore cannot be extrapolated to substantially smaller drift using the generic linear law. A strongly directed flow also has an approximate quadratic torque dependence for a different reason: its incident momentum flux is proportional to $v_{\rm drift}^2$.

For gas-damped grains, the definitions in Section~2 give $\tau_{\rm drag}\propto a$, $I_1\propto a^5$, and $J_{\rm th}\propto a^{5/2}$ at fixed gas conditions. At a stable high-$J$ equilibrium, the weak-drift scaling gives
\begin{equation}
 \frac{|J_*|}{J_{\rm th}}\propto a^{3/2}C_Hs_d,
 \qquad \omega_*\propto \frac{C_Hv_{\rm drift}}{a}.
 \label{eq:met-simple-spin}
\end{equation}
Thus decreasing $s_d$ does not impose a sharp cutoff in MET driving: whether grains rotate suprathermally depends on the product $a^{3/2}C_Hs_d$. The coefficient is given in Eq.~\eqref{eq:disk-met-spin}. Extra rotational damping processes can reduce $|J_*|$, especially for the smallest grains.

\section{Shocks, Ambipolar Drift, and Radiation Pressure}\label{sec:drift-examples}
\subsection{Shocked regions}\label{sec:shock}
In magnetized shocks, the relevant MET driver is the local grain--neutral slip, not the shock speed. Its magnitude and angle $\beta$ depend on grain charge, size, and coupling to the field \citep{ChapmanWardle2006}. Oblique drift can produce stable high-$J$ alignment even with weak magnetic relaxation (Figure~\ref{fig:scan}).

Grains driven toward a low-$J$ attractor can remain trapped during the shock passage. Gas impacts can displace grains across the low-$J$ basin boundary, as in RAT alignment \citep{HoangLazarian2008,LazarianHoang2021Disruption}. In fast drift, strong bombardment can also deflect an escaping grain before coherent MET spin-up carries it to high $J$. Repeated recapture can delay arrival at a high-$J$ attractor and delay centrifugal disruption on trajectories toward either a high-$J$ attractor or a high-$J$ repeller (Section~\ref{sec:met-disruption}). The relevant times are the first successful arrival at high $J$ and the first crossing of $J_{\rm disr}$, respectively. Both can exceed the time of the first departure from low $J$.

A grain crossing a $10^{16}$ cm shock layer at $2\,\mathrm{km\,s^{-1}}$ in the shock frame spends about $1.6\times10^3$ yr in the layer. For $n_{\rm H_2}=10^4\,\mathrm{cm^{-3}}$, $T_g=100$ K, and a compact grain of radius $0.03\,\mu$m and density $3\,\mathrm{g\,cm^{-3}}$, this is about $24\tau_{\rm drag}$ using the thermal gas-damping normalization in Section~2. Grains driven directly toward a high-$J$ attractor can align during passage. For grains initially trapped at low $J$, the fraction that reaches a high-$J$ attractor during passage is the cumulative probability of successful first arrival evaluated at the layer-crossing time. The disrupted fraction instead depends on the cumulative probability of reaching $J_{\rm disr}$ during passage.

For very small grains in fast shocks, stochastic impacts can sustain higher rotation rates than coherent METs (Eq.~\ref{eq:met-coherent-stochastic}). Anisotropic bombardment by a supersonic grain--neutral flow can then align the grains through the Gold mechanism \citep{Gold1952,Lazarian1994}. Surviving nanoparticles with a permanent magnetic moment can contribute polarized microwave emission as they spin \citep{HoangLazarian2016SpinningMagnetic}. For principal-axis rotation, the time-varying magnetic dipole moment lies in the plane perpendicular to $\mathbf J$. Its emission has an electric vector preferentially parallel to the projection of $\mathbf J$ on the sky. The polarization can therefore be parallel to the projected shock normal when Gold alignment preferentially orients $\mathbf J$ along the normal. This geometry is possible when rapid magnetic precession about a field along the normal averages the impacts from a predominantly transverse slip \citep[cf.][]{RobergeHananyMessinger1995}. With predominantly normal slip and negligible magnetic precession, Gold alignment instead favors $\mathbf J$ in the shock plane and the predicted polarization is perpendicular to the projected normal.

\subsection{Dark Clouds}\label{sec:ambipolar}
Ambipolar diffusion causes charged material to slip relative to neutral gas. Gold-type alignment driven by ambipolar diffusion was studied by \citet{RobergeHananyMessinger1995,DeschRoberge1997}. For irregular grains, coherent METs can provide systematic high-$J$ spin-up and alignment rather than stochastic Gold alignment. The ion--neutral velocity difference obeys \citep{RobergeHananyMessinger1995,DeschRoberge1997}
\begin{equation}
 \gamma\rho_i\rho_n(\mathbf v_i-\mathbf v_n)\simeq
 \frac{(\boldsymbol\nabla\times\mathbf B)\times\mathbf B}{4\pi},
 \label{eq:adbalance}
\end{equation}
when ion inertia is negligible. The grain--neutral slip is generally smaller because grains respond to both Lorentz forces and gas drag. Writing $\omega_g=Z_geB/(m_{\rm gr}c)$, the grain Hall parameter is \citep{RobergeHananyMessinger1995,GuilletEtAl2020}
\begin{equation}
 \mathcal H_g\simeq\frac{|Z_g|eB}{m_{\rm gr}c}\,\tau_{\rm drag}
 \simeq\frac{3|Z_g|eB}{4\pi c\,a^2\rho_n v_{\rm th}},
 \label{eq:grhall}
\end{equation}
where the second expression assumes a compact spherical grain and a drag time of order $\rho a/(\rho_n v_{\rm th})$. For a steady ion--neutral drift perpendicular to $\mathbf B$, the grain--neutral slip is \citep{GuilletEtAl2020}
\begin{equation}
 v_{\rm drift}(a)\simeq v_{\rm AD}\,
 \frac{\mathcal H_g(a)}{\sqrt{1+\mathcal H_g^2(a)}},
 \qquad v_{\rm AD}=|\mathbf v_i-\mathbf v_n|.
 \label{eq:adgrainslip}
\end{equation}
The drift remains perpendicular to the local field in this steady model. At fixed charge, $\mathcal H_g\propto a^{-2}$, so small charged grains are better coupled to the ions than large grains.

At $T_g=10$ K the molecular-gas thermal speed is about $0.3\,\mathrm{km\,s^{-1}}$. For $n_{\rm H_2}=10^4\,\mathrm{cm^{-3}}$, $B=30\,\mu\mathrm G$, $Z_g=-1$, and grain density $2.2\,\mathrm{g\,cm^{-3}}$, the Hall parameters at $a=0.03$, 0.1, and $0.3\,\mu$m are approximately 12, 1.1, and 0.12. Even if the ion--neutral speed is $v_{\rm AD}=0.3\,\mathrm{km\,s^{-1}}$, the corresponding grain drift parameters are only $s_d\simeq1.1$, 0.84, and 0.14. Thus grain--neutral drift, rather than ion--neutral drift, must enter the MET torque calculation.

Table~\ref{tab:cloud-drift-delta} shows a weaker steady drift, $v_{\rm AD}=0.10\,\mathrm{km\,s^{-1}}$, together with the magnetic-relaxation parameter for two magnetic-response strengths. The numerical MET survey at $s_d=0.3$ is more representative of these cloud parameters than the $s_d=1$ calculation. The enhanced response $K=10^{-11}$ s gives $\delta_m\simeq5$ for $0.03\,\mu$m grains and 1.5 for $0.1\,\mu$m grains. The drift simultaneously weakens toward larger sizes.

\begin{table*}[t]
\centering
\caption{Steady grain--neutral drift and magnetic relaxation versus grain radius in an illustrative dark cloud.}\label{tab:cloud-drift-delta}
\small
\begin{tabular}{cccccc}
\hline\hline
$a\ (\mu{\rm m})$ & $\mathcal H_g$ & $v_{\rm drift}\ ({\rm m\,s^{-1}})$ & $s_d$ & $K\ ({\rm s})$ & $\delta_m$ \\
\hline
0.01 & 110 & 100 & 0.4 & $10^{-13}$ & 0.15 \\
     &     &     &       & $10^{-11}$ & 15 \\
0.03 & 12 & 100 & 0.4 & $10^{-13}$ & 0.05 \\
     &      &      &       & $10^{-11}$ & 5 \\
0.10 & 1.1 & 74 & 0.3 & $10^{-13}$ & 0.015 \\
     &      &      &       & $10^{-11}$ & 1.5 \\
0.30 & 0.12 & 12 & 0.05 & $10^{-13}$ & 0.005 \\
     &       &      &       & $10^{-11}$ & 0.5 \\
\hline
\end{tabular}
\par\smallskip
\begin{minipage}{0.88\textwidth}\footnotesize
\emph{Note.} Illustrative steady, single-charge model with $n_{\rm H_2}=10^4\,{\rm cm^{-3}}$, $T_g=10\,$K, $B=30\,\mu{\rm G}$, $Z_g=-1$, and $v_{\rm AD}=0.10\,{\rm km\,s^{-1}}$. Each grain radius has two rows showing the same $\delta_m$ evaluated for the weak ($K=10^{-13}\,{\rm s}$) and enhanced ($K=10^{-11}\,{\rm s}$) magnetic responses. Blank entries repeat the grain properties in the row above.
\end{minipage}
\end{table*}

The steady ambipolar slip is transverse to $\mathbf B$, but turbulent grain motions and finite-inertia response to a wandering field can make the grain--neutral drift oblique \citep{YanLazarianDraine2004,LazarianVishniac1999}:
\begin{equation}
 \mathbf v_{\rm drift}=\mathbf v_{\rm AD,gr}+\delta\mathbf v_{\rm turb},
 \qquad
 \cos\beta=\frac{\mathbf v_{\rm drift}\cdot\mathbf B}{|\mathbf v_{\rm drift}|B},
 \label{eq:ad-turb-beta}
\end{equation}
For example, $|v_\parallel|=v_\perp|\cot\beta|$ requires only about $15\,\mathrm{m\,s^{-1}}$ along the field for $v_\perp=0.84\,\mathrm{km\,s^{-1}}$ and $\beta=89^\circ$. The ideal ion--neutral slip remains transverse to the instantaneous field, even when grain inertia makes the grain--neutral velocity oblique.

Direct calculations confirm the importance of this geometry. At $\beta=90^\circ$ and $\delta_m=2$, the saturated mixed-law $R_B$ rises from roughly 0.6 at subsonic drift to nearly unity at stronger drift. Enhanced magnetic relaxation also removes much of the sharp sensitivity to near-transverse angles found at $\delta_m=1$. The first-order magnetic-axis spin-up nevertheless vanishes at exactly $90^\circ$ in the weak-drift limit (Eq.~\eqref{eq:met-weak-transverse}).

The grain-size dependence combines $\mathcal H_g\propto a^{-2}$ with the normalized MET spin-up proportional to $a^{3/2}$ when the linear drift coefficient is nonzero. For the tabulated cloud parameters, reaching $|J_*|\gtrsim3J_{\rm th}$ requires $C_H\equiv|H_B|/s_d\gtrsim3\times10^{-3}$ at $0.03\,\mu$m or $6\times10^{-4}$ at $0.1\,\mu$m, together with a stable high-$J$ attractor. In shielded cores, this provides an independent potential source of alignment when RATs weaken. The decline of polarization toward the center of Pipe-109 \citep{AlvesEtAl2014Pipe109} is consistent with regions where grain--gas slip is too small to sustain efficient MET alignment. A direct test compares polarization with independently estimated grain drift and magnetic geometry.

\subsection{Circumstellar regions}\label{sec:circumstellar-pressure}
Radiation pressure can accelerate dust relative to circumstellar gas. Balancing the radiative force with supersonic free-molecular drag gives the indicative terminal speed
\begin{align}
 v_{\rm drift}&\sim
 \left(\frac{Q_{\rm pr}L_*}{4\pi r^2c\rho_g}\right)^{1/2}
 \nonumber\\
 &\simeq16.5\,\mathrm{km\,s^{-1}}
 \left(\frac{Q_{\rm pr}}{0.1}\right)^{1/2}
 \left(\frac{L_*}{10^4L_\odot}\right)^{1/2}
 \nonumber\\
 &\quad\times\left(\frac{r}{100\,\mathrm{au}}\right)^{-1}
 \left(\frac{n_g}{10^6\,\mathrm{cm^{-3}}}\right)^{-1/2}
 \nonumber\\
 &\quad\times\left(\frac{m_g}{m_{\rm H}}\right)^{-1/2}.
 \label{eq:radiation-pressure-drift}
\end{align}
where $Q_{\rm pr}$ is the pressure efficiency, $L_*$ is the stellar luminosity, $r$ is the source distance, and $\rho_g=n_gm_g$. Luminous stars can therefore sustain supersonic grain--gas drift.

Irregular PAH-sized particles and ultrasmall silicate grains could experience mechanical spin-up under this drift, with implications for polarized spinning-dust emission \citep{DraineLazarian1998Spinning,HoangVinhLan2016}. The $a^{3/2}$ scaling makes high-$J$ rotation harder for nanoparticles. For a 1 nm grain with $\rho=2.2\,\mathrm{g\,cm^{-3}}$, the weak-drift estimate requires $C_Hs_d\gtrsim0.17$ for $|j_*|\ge3J_{\rm th}$. At $s_d=0.3$, this corresponds to $C_H\gtrsim0.6$ at 1 nm and 0.05 at 5 nm. Symmetric PAHs have little systematic MET, although radiative cross-sectional alignment offers a separate route to polarized spinning dust \citep{HoangLazarian2018CrossSection}.

For larger grains, mechanical precession can outpace Larmor precession and select the drift direction as the alignment axis. Radial drift and a radial magnetic field in a stellar wind then lead to coincident axes, whereas a spiral magnetic field produces a changing geometry with distance. Strong drift can also produce rotational disruption (Section~\ref{sec:met-disruption}).

\section{MET Alignment in Circumstellar Disks}\label{sec:disk-applications}

METs were previously investigated in protostellar disks by \citet{HoangEtAl2022Disk}. Our calculation tests how the stability of high-$J$ alignment depends on grain geometry, magnetic relaxation, and drift direction. Two quantities determine its disk interpretation: the axis about which the grain precesses and the relative grain--gas velocity.

\subsection{Alignment axis and polarization}\label{sec:disk-axis}

When Larmor precession dominates, stable high-$J$ grains align with the magnetic field. When mechanical precession dominates, the preferred axis is the grain--gas drift, provided the METs can spin up and align the grain. For a face-on disk with a toroidal field and predominantly radial drift, the two cases give different thermal-emission patterns: radial polarization for $B$-alignment and azimuthal polarization for drift alignment. Azimuthal drift at ${\rm St}\sim1$ or vertical settling can modify these patterns \citep{KataokaOkuzumiTazaki2019}. The actual polarization also depends on internal alignment and optical depth.

Observed radial polarization in the southern part of HD~142527 is consistent with emission tracing a toroidal field, while longer-wavelength azimuthal polarization in HL~Tau and other disk regions motivates considering drift-axis alignment \citep{OhashiEtAl2025HD142527,StephensEtAl2023HLTau,LinEtAl2024HLTau}. Scattering and RAT alignment remain alternative contributors. A MET interpretation predicts a change in polarization direction where the dominant precession axis or the grain--gas drift changes.

\subsection{Disk drift and mechanical spin-up}\label{sec:disk-drift-estimate}

In a pressure-supported disk, grains drift relative to the gas both radially and through vertical settling. For Epstein drag and $\mathrm{St}\ll1$, we use \citep{BirnstielFangJohansen2016,Birnstiel2024}
\begin{align}
 \mathrm{St}(a,z)&\simeq \frac{\pi\rho a}{2\Sigma_g}\exp\!\left(\frac{z^2}{2H_g^2}\right),\quad H_g=\frac{c_s}{\Omega_K},
 \label{eq:disk-stokes}\\
 v_{r,\rm drift}&\simeq-2\,\mathrm{St}\,\eta v_K,\qquad
 v_{z,\rm drift}\simeq-\mathrm{St}\,\Omega_Kz,
 \label{eq:disk-drift-rz}
\end{align}
where $\eta v_K$ is the gas pressure-support speed. For larger stopping times the azimuthal drift matters, with the standard pressure-driven solution
\begin{equation}
 \Delta v_r\simeq-\frac{2\eta v_K\,\mathrm{St}}{1+\mathrm{St}^2},\qquad
 \Delta v_\phi\simeq\frac{\eta v_K\,\mathrm{St}^2}{1+\mathrm{St}^2}.
 \label{eq:disk-drift-rphi}
\end{equation}

\begin{figure}[t]
\centering
\includegraphics[width=\columnwidth]{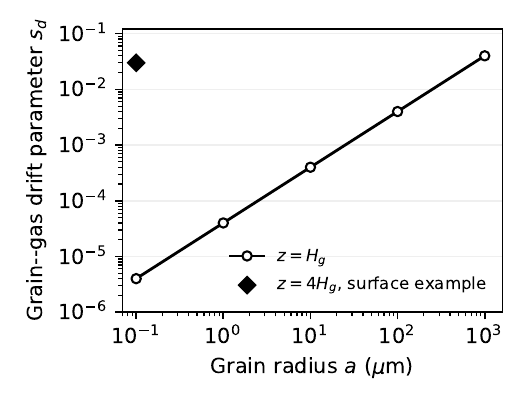}
\caption{Grain--gas drift parameter for the illustrative disk at 30 au. The curve shows $z=H_g$; the diamond gives the separate surface-layer example at $z=4H_g$.}
\label{fig:disk-compact}
\end{figure}

For illustration, we take a solar-mass star at 30 au, $T_g=30\,$K, $\Sigma_g=10\,\mathrm{g\,cm^{-2}}$, $\eta v_K=40\,\mathrm{m\,s^{-1}}$, and grain density $\rho=2.2\,\mathrm{g\,cm^{-3}}$. At $z=H_g\simeq1.8$ au, the predicted drift increases from $s_d\sim4\times10^{-6}$ for $a=0.1\,\mu$m to $s_d\sim0.04$ at 1 mm (Figure~\ref{fig:disk-compact}). This is subsonic drift, so the weak-drift MET scaling is appropriate.

Define the effective torque slope $C_H=|H_B|/s_d$ for an individual grain in the weak-drift limit. Using the reference damping time and inertia of Section~2, its equilibrium spin is
\begin{equation}
 \frac{J_{\rm MET}}{J_{\rm th}}\simeq
 \frac{0.7}{\sqrt{8\pi/15}}\sqrt{\frac{2\rho}{m_g}}\,a^{3/2}C_Hs_d,
 \label{eq:disk-met-spin}
\end{equation}
provided a stable spin-up orientation exists. In the example above, reaching $3J_{\rm th}$ requires $C_H\sim0.1$ for $1\,\mu$m grains, $4\times10^{-4}$ at $10\,\mu$m, and $10^{-6}$ at $100\,\mu$m. Since $s_d\propto a$ at fixed location for $\mathrm{St}\ll1$, the spin-up scales approximately as $a^{5/2}$ for similar torque geometry. Large grains can therefore experience strong MET spin-up even at subsonic drift.

In a shielded layer with $\bar\lambda\sim100\,\mu$m and $u_{\rm rad}\sim6\times10^{-9}\,\mathrm{erg\,cm^{-3}}$, the illustrative choice $C_H=0.02$ gives $J_{\rm MET}/J_{\rm th}\sim150$ and $\Gamma_{\rm MET}/\Gamma_{\rm RAT}\sim400$ for $10\,\mu$m grains, but only $J_{\rm MET}/J_{\rm th}\sim0.5$ at $1\,\mu$m. These values compare torque strengths, not the fraction of grains captured at high $J$.

A disk surface can behave differently from the denser layer. At $z=4H_g$, our adopted field $B\simeq12\,$mG, $s_d\sim0.03$, and ordinary paramagnetic response $K=6\times10^{-14}\,$s yield a saturated magnetic-axis factor $R_B\sim0.8$ for $0.1\,\mu$m grains, at drift--field angle $75^\circ$. This is an illustrative strongly magnetized, tenuous layer; it need not dominate the disk's polarized emission. 

\subsection{Grain growth and mantle loss}\label{sec:disk-pebble-growth}

MET spin-up may remove icy mantles from grains that reach stable high-$J$ rotation. Equating centrifugal stress to an effective ice adhesion strength gives
\begin{equation}
 \omega_{\rm ice}\simeq\left[\frac{4S_{\rm ice}}{\rho_{\rm ice}(a^2-a_c^2)}\right]^{1/2},
 \label{eq:met-disk-mantle-loss}
\end{equation}
where $a_c$ is the core radius. Grains remaining at low $J$ can retain their mantles. Spin-up can also supply frictional energy during grain collisions, alongside the anisotropic growth of aligned grains discussed by \citet{Hoang2022AlignedGrowth,HoangEtAl2022Disk}. The outcome depends on high-$J$ occupation and exposure time, not spin-up efficiency alone.

\subsection{Comparison with earlier MET studies of disks}\label{sec:disk-prior-work}

\citet{HoangEtAl2022Disk} established that METs can compete with or exceed RATs in protostellar disks and examined internal relaxation and the competition between magnetic-field and drift-axis alignment. Our calculation complements that work by deriving torque functions for irregular shapes, identifying stable high-$J$ attractors, and calculating their alignment factors rather than assigning representative MET efficiencies. We also include radial and settling drift explicitly. Our numerical factors assume effective internal alignment of grains that reach the high-$J$ states; the earlier internal-relaxation constraints remain relevant for very large grains.

\section{Rotational disruption by mechanical torques}\label{sec:met-disruption}

METs can spin irregular grains above the centrifugal disruption threshold, as RATs do \citep{HoangEtAl2019RATD,LazarianHoang2021Disruption}. For reference material density $\rho$, radius $a$, and strength $S_{\max}$, we use \citep{Hoang2019GrainModels,Hoang2020DisruptionReview}
\begin{equation}
 \omega_{\rm disr}\simeq\frac{2}{a}\left(\frac{S_{\max}}{\rho}\right)^{1/2},
 \qquad J_{\rm disr}=I_1\omega_{\rm disr}.
 \label{eq:met-disr-threshold}
\end{equation}
We compare this reference threshold with the calculated equilibrium spins of the grain sample. Their actual local failure stresses have not been modeled. For a stable high-$J$ state at inclination $\psi_*$,
\begin{equation}
 \frac{|J_*|}{J_{\rm disr}}=
 \frac{\Gamma_0\tau_{\rm drag}|H_B(\psi_*)|}
 {I_1\omega_{\rm disr}[1+\delta_m\sin^2\psi_*]}.
 \label{eq:met-disr-ratio}
\end{equation}
Along a spin-up trajectory toward such a state, the idealized threshold-crossing time is
\begin{equation}
 \frac{t_{\rm disr}}{\tau_{\rm eff}}=
 -\ln\left(1-\frac{J_{\rm disr}}{|J_*|}\right),\qquad |J_*|>J_{\rm disr}.
 \label{eq:met-disr-time}
\end{equation}
This expression assumes a coherent drift episode and spin-up along the stable branch. It gives $t_{\rm disr}\sim0.7\tau_{\rm eff}$ when $|J_*|/J_{\rm disr}=2$ and $\sim0.2\tau_{\rm eff}$ when the ratio is 5. For grains initially trapped at low $J$, the time to enter the high-$J$ basin can dominate the total disruption time. Disruption can also occur during transient spin-up near a high-$J$ repeller if the grain crosses $J_{\rm disr}$ before the angular dynamics carry it away \citep{LazarianHoang2021Disruption}. Low-$J$ trapping can delay access to either route.

Motivated by the grain--gas drift scaling in the driven-turbulence simulations of \citet{MoseleyTeyssier2025}, we use
\begin{equation}
 v_{\rm drift,rms}\simeq0.5\left(\frac{a}{0.1\,\mu\mathrm m}\right)^{1/2}\mathrm{km\,s^{-1}}
 \label{eq:met-turb-drift}
\end{equation}
for a representative cold neutral medium, with $T_g=100\,$K, $n_{\rm H}=30\,\mathrm{cm^{-3}}$, $B=5\,\mu$G, and $\rho=3\,\mathrm{g\,cm^{-3}}$. The chosen rms velocity is treated as a quasi-steady drift only during the spin-up episode. Two material models are compared: weak aggregates ($S_{\max}=10^6\,\mathrm{erg\,cm^{-3}}$, $K=10^{-14}\,$s) and compact silicates ($S_{\max}=10^{10}\,\mathrm{erg\,cm^{-3}}$, $K=10^{-13}\,$s).

\begin{figure}[t]
\centering
\includegraphics[width=\columnwidth]{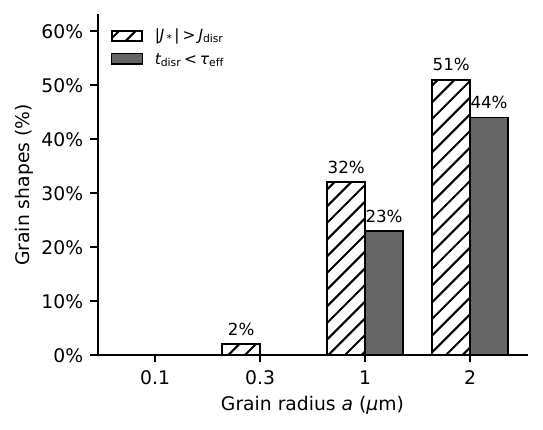}
\caption{Percentage of weak-aggregate grain shapes with a stable high-$J$ equilibrium spin exceeding the centrifugal disruption threshold (hatched), and percentage capable of reaching disruption within one effective damping time along the assumed spin-up trajectory (gray). Results are for $\beta=75^\circ$ and the grain-size-dependent drift of Eq.~\eqref{eq:met-turb-drift}. No compact silicate in the sample disrupts under these assumptions.}
\label{fig:met-disruption-compact}
\end{figure}

Figure~\ref{fig:met-disruption-compact} shows the results at drift--field angle $\beta=75^\circ$. For weak aggregates, the fraction of grains with $|J_*|>J_{\rm disr}$ rises from $2\%$ at $0.3\,\mu$m to $32\%$ at $1\,\mu$m and $51\%$ at $2\,\mu$m. The corresponding fractions that can reach disruption within one effective damping time are $0\%$, $23\%$, and $44\%$. Both fractions vanish at $0.1\,\mu$m. None of the compact silicates in this calculation crosses its much higher threshold. Exactly transverse drift can also suppress qualifying stable high-$J$ states at weak magnetic relaxation. These fractions apply to the stable-high-$J$ channel under the adopted steady drift. They are not the overall destruction fractions in time-dependent turbulence.

\citet{HoangLee2020METD} investigated disruption by stochastic gas impacts in fast shocks and radiation-driven flows, including nanometer-sized grains. For the directed, shape-dependent METs considered here, a characteristic angular momentum $J_{\rm sys}=\epsilon_{\rm spin}I_1v_{\rm drift}/a$ gives the indicative comparison
\begin{equation}
 \frac{J_{\rm sys}}{J_{\rm stoch}}\sim
 \epsilon_{\rm spin}\left(\frac{m_{\rm gr}}{m_{\rm gas}}\right)^{1/2},\qquad
 m_{\rm gr}\sim\frac{4\pi}{3}\rho a^3.
 \label{eq:met-coherent-stochastic}
\end{equation}
Thus coherent torques become more important relative to random impacts as grain size increases, provided drift and grain orientation remain correlated. At $v_{\rm drift}=100\,\mathrm{km\,s^{-1}}$, their centrifugal significance is approximately
\begin{equation}
 \frac{J_{\rm sys}}{J_{\rm disr}}\sim
 \frac{\epsilon_{\rm spin}v_{\rm drift}}
 {2\sqrt{S_{\max}/\rho}}.
 \label{eq:met-fast-drift-breakup}
\end{equation}
For $\epsilon_{\rm spin}\sim10^{-3}$--$10^{-2}$, this ratio is of order $0.3$--3 for $S_{\max}=10^9\,\mathrm{erg\,cm^{-3}}$, and $10$--100 for $S_{\max}=10^6\,\mathrm{erg\,cm^{-3}}$. Random impacts can disrupt very small grains, while systematic METs can become effective for sufficiently irregular larger grains. The present ensemble calculations quantify only the coherent, stable-high-$J$ route. Disruption preferentially removes susceptible high-$J$ grains and may thereby alter the polarized grain population.

\section{Discussion}
\subsection{Composition-dependent and wavelength-dependent polarization}\label{sec:composition-polarization}
Different compositions and sizes can select different reference axes. Paramagnetic silicates may follow $\mathbf B$, whereas carbonaceous grains with weak magnetic response may precess about their relative drift \citep{Lazarian2020CarbonModes}. In ambipolar diffusion, where drift is perpendicular to $\mathbf B$, the two populations can produce thermal-emission polarization directions separated by $90^\circ$ if the projected axes remain perpendicular. In shocks the separation can take other values. These predictions require stable drift-axis rotation for the weakly magnetic grains. The present numerical $R_B$ calculations assume fast $B$ precession.

The disk polarization patterns for magnetic-field and drift-axis MET alignment are discussed in Section~\ref{sec:disk-axis}. The measured polarization also contains contributions from RAT alignment and dust scattering \citep{HoangEtAl2022Disk}.

\subsection{Anisotropic growth and frictional bonding}\label{sec:met-coagulation}
\citet{Hoang2022AlignedGrowth} showed that the preferred orientation of RAT-aligned grains can make their coagulation anisotropic. \citet{HoangEtAl2022Disk} also discussed anisotropic aggregation of MET-aligned grains. These studies concern how grain alignment controls collision geometry and the direction of growth. We propose an additional process: \emph{MET-powered frictional bonding}, in which the rotational energy of grains is dissipated at the contact during a collision.

The physical analogy is rotary friction welding, where a rotating workpiece is pressed against another and interfacial sliding generates heat and deformation that create a bond \citep{LiEtAl2016FrictionWelding}. Astrophysical grain collisions supply the contact force for a much shorter time, but grains spun up by METs carry rotational energy into the encounter. If two irregular surfaces slide against one another, tangential friction can dissipate that energy and strengthen adhesion. Rolling and sliding friction are already known to dissipate energy in dust-aggregate contacts \citep{DominikTielens1995Rolling}. For ice-coated grains, local restructuring and heating can further enhance bonding \citep{NietiadiEtAl2017IceMelting}. The proposed process differs from Hoang's anisotropic-growth mechanism: alignment determines the preferred collision geometry, whereas MET-driven rotation provides an additional energy source for contact bonding.

The surface speed is $v_{\rm rot}=a\omega$. Using the spin efficiency $\epsilon_{\rm spin}$ defined in Section~\ref{sec:met-disruption}, we obtain $v_{\rm rot}\sim\epsilon_{\rm spin}v_{\rm drift}\sim0.1$--$1\,{\rm m\,s^{-1}}$ for $v_{\rm drift}=100\,{\rm m\,s^{-1}}$ and $\epsilon_{\rm spin}=10^{-3}$--$10^{-2}$. For two equal approximately spherical grains, each of mass $m$, the combined spin energy is $E_{\rm rot}\sim(2/5)m v_{\rm rot}^2$, compared with the relative collision energy $E_{\rm col}=m v_{\rm col}^2/4$. Frictional loss of even part of the spin energy can change whether the grains rebound or remain bound. Contact bonding is particularly plausible for grains bearing ice or organic mantles. At greater rotational or collision speeds, fragmentation and centrifugal disruption compete with coagulation \citep{TatsuumaKataoka2021GasFlow}.

The two processes can reinforce each other in dark clouds and circumstellar disks. For grains whose spin axes are aligned with $\mathbf B$, rotational surface motion occurs predominantly perpendicular to the field. The orientation-dependent collisions discussed by \citet{Hoang2022AlignedGrowth} may then preferentially extend aggregates perpendicular to $\mathbf B$, while frictional dissipation strengthens the newly formed contacts. The sliding is caused by grain rotation about its own axis, not by Larmor precession. METs can replenish rotational energy between collisions, potentially influencing the growth of large aggregates on the way to planetesimals.

\subsection{Synergy with RATs}\label{sec:met-rat-synergy}
METs provide a complementary alignment mechanism where RATs are inefficient: in shielded regions with weak or attenuated radiation and for grain sizes poorly coupled to the available radiation wavelengths. Grain--gas drift supplies an independent source of anisotropic forcing, so METs can maintain alignment when radiative spin-up is insufficient. The relative torque strengths and grain-size regimes are quantified in Section~\ref{sec:RATcompare}.

A second possibility is that METs determine the \emph{alignment axis} even when RATs supply most of the aligning torque and spin-up. The sizable mechanical $Q_{e3}$ quantified above can drive rapid precession about the drift direction. If this precession dominates radiative and Larmor precession, the RAT aligning torque is averaged about the drift axis, and its stable attractors are defined relative to that axis rather than automatically to the radiation direction or magnetic field \citep{LH2007rat,LH2007met}. Thus METs can contribute to alignment either directly, through their orienting torque, or indirectly, by selecting the axis about which RAT alignment operates.

\section{Summary}
We calculated regular mechanical torques for the grain sample using the adopted collision law and checked the two pure-collision limits. We evaluate alignment from stable high-$J$ states, with no low-$J$ contribution, and compare its magnetic enhancement with the RAT framework. Our main findings are:
\begin{enumerate}
\item The ensemble-mean $Q_{e1}$ and $Q_{e2}$ angular profiles are AMO-like even though individual MET profiles vary strongly with grain shape, collision law, and drift. The original median $q_{\max}$ which is ratio of the amplitudes of the $Q_{e1}$ and $Q_{e2}$ torques, decreases over the sampled drift range. An independent grain-shape calculation confirms the predicted individual-grain weak-drift limit $q_{\max}\to 4/3$, supporting analytical MET descriptions.
\item Paramagnetic relaxation enhance oblique-drift alignment. Enhanced magnetic response of grains makes the alignment perfect in respect to magnetic field.
\item Ambipolar diffusion, shocks, radiation pressure and turbulence provide distinct sources of grain--gas drift. METs can produce composition- and wavelength-dependent polarization, as well as to influence the grain-size distribution by inducing disruption and coagulation of grains. 
\item Compared with RATs, METs can align grains in weak or strongly attenuated radiation fields. Their relative torque efficiency also increases for grains much smaller or much larger than the characteristic radiation wavelength: the long-wavelength RAT efficiency falls as $(a/\bar\lambda)^3$, while for $a\gg\bar\lambda$ the METs can dominate the alignment. METs can provide alignment nanograins inducing polarization of their microwave emission.
\item METs are particular important for accretion disks, where they potentially can explain some features of observed polarization as well as provide an insight into the evolution of grains and their icy mantles.
\item METs can control the grain discuption and growth in accretion disks at depths and grain sizes for which RATs are not efficient.
\item A process analogous of frictional welding was identified for coagulation of aligned grains.

\end{enumerate}


\appendix
\section{Magnetic-field precession averages of the MET functions}\label{app:B-average}

We give the explicit relation between the torque components defined relative to the grain--gas drift and the functions averaged about a rapidly precessed magnetic axis. This also shows why $\beta=0$ is the drift-axis limit. We use $\hat{\mathbf u}\parallel\mathbf J$ for positive signed spin, $\hat{\mathbf k}$ along the directed drift, $\psi=\angle(\hat{\mathbf u},\hat{\mathbf B})$, $\beta=\angle(\hat{\mathbf k},\hat{\mathbf B})$, and $\Theta=\angle(\hat{\mathbf u},\hat{\mathbf k})$. The signed-$j$ convention in the main text is retained.

Choose the magnetic field as the polar axis, with the drift in the zero-azimuth plane. At magnetic precession phase $\varphi$,
\begin{equation}
 c\equiv\cos\Theta=\cos\psi\cos\beta+\sin\psi\sin\beta\cos\varphi.
 \label{eq:app-angle}
\end{equation}
The spin-averaged scattering-frame torques are
\begin{align}
 H(\Theta)&=Q_{e1}(\Theta)\cos\Theta+Q_{e2}(\Theta)\sin\Theta,\\
 F(\Theta)&=-Q_{e1}(\Theta)\sin\Theta+Q_{e2}(\Theta)\cos\Theta,\\
 G(\Theta)&=Q_{e3}(\Theta).
 \label{eq:app-HFG}
\end{align}
Writing $\langle\cdot\rangle_\varphi=(2\pi)^{-1}\int_0^{2\pi}d\varphi(\cdot)$, direct vector projection gives
\begin{align}
 H_B(\psi;\beta)&=\left\langle Q_{e1}(\Theta)c+Q_{e2}(\Theta)\sqrt{1-c^2}\right\rangle_\varphi,\\
 F_B(\psi;\beta)&=\left\langle\left[-Q_{e1}(\Theta)\sqrt{1-c^2}+Q_{e2}(\Theta)c\right]
 \frac{\cos\beta-\cos\psi\,c}{\sin\psi\sqrt{1-c^2}}\right\rangle_\varphi.
 \label{eq:app-B-general}
\end{align}
These are the magnetic averages used in the main text. The projection of $G$ onto the magnetic-polar direction is odd in $\varphi$ and integrates to zero. Continuous angular limits are understood when $\sin\Theta$ or $\sin\psi$ vanishes.

\subsection{Exact aligned-axis limit}\label{app:beta-zero}
For $\beta=0$, $\hat{\mathbf B}=\hat{\mathbf k}$ and Eq.~\eqref{eq:app-angle} reduces to $\Theta=\psi$ independently of precession phase. The projection factor in Eq.~\eqref{eq:app-B-general} is one, and therefore
\begin{equation}
 H_B(\psi;0)=H(\psi)=H_k(\psi),\qquad
 F_B(\psi;0)=F(\psi)=F_k(\psi).
 \label{eq:app-identity}
\end{equation}
This is an identity of the averaged torque functions, not an approximation at small drift speed. The two descriptions are the same dynamical problem when the reference axes coincide. In particular, at $\delta_m=0$ the sample $R_B$ evaluated at $\beta=0$ is already its drift-axis Rayleigh factor. A change in the source of the rapid precession cannot change this aligned-axis angular torque. If magnetic relaxation is retained, its drift-axis and magnetic-axis contributions also coincide at $\beta=0$:
\begin{equation}
 \left.\dot\psi\right|_{\rm mag}=-\frac{\delta_m}{\tau_{\rm drag}}\sin\psi\cos\psi,
 \qquad
 \left.\dot j\right|_{\rm mag}=-\frac{\delta_m j}{\tau_{\rm drag}}\sin^2\psi.
\end{equation}

For comparison, averaging magnetic relaxation about a \emph{different}, rapidly precessed drift axis gives
\begin{align}
 \left\langle\sin^2\psi_B\right\rangle_{\varphi_k}
 &=1-\cos^2\Theta\cos^2\beta-\tfrac12\sin^2\Theta\sin^2\beta,\\
 \left.\dot\Theta\right|_{\rm mag,k}
 &=-\frac{\delta_m}{\tau_{\rm drag}}P_2(\cos\beta)\sin\Theta\cos\Theta.
 \label{eq:app-k-magnetic}
\end{align}
The first expression determines the spin damping and the second the polar relaxation. Both reduce to the usual magnetic-axis terms at $\beta=0$.

\subsection{Analytical averages of the lowest-order angular model}
The AMO-like illustration in Eq.~\eqref{eq:qmax-amo-template} permits closed-form magnetic averages. Let $q=q_{\max}$ and define
\begin{equation}
 A=2-\frac{2q}{3},\qquad B=\frac{5q}{3}-2,\qquad C=\frac{2q}{3}.
\end{equation}
The unaveraged torques are then $H(\Theta)=c(A+Bc^2)$ and $F(\Theta)=\sin\Theta(C-Bc^2)$. Introduce $x=\cos\psi$ and $y=\cos\beta$. Using
\begin{equation}
 \langle c^2\rangle_\varphi=x^2y^2+\tfrac12(1-x^2)(1-y^2),\quad
 \langle c^3\rangle_\varphi=x^3y^3+\tfrac32xy(1-x^2)(1-y^2),
\end{equation}
we obtain
\begin{align}
 H_B(\psi;\beta)&=xy\left[A+B\left(x^2y^2+\tfrac32(1-x^2)(1-y^2)\right)\right],
 \label{eq:app-HB-amo}\\
 F_B(\psi;\beta)&=y\sin\psi\left[C-B\left(x^2y^2+\tfrac12(1-3x^2)(1-y^2)\right)\right].
 \label{eq:app-FB-amo}
\end{align}
For $y=1$ these expressions become $H_B=x(A+Bx^2)=H(\psi)$ and $F_B=\sin\psi(C-Bx^2)=F(\psi)$, verifying Eq.~\eqref{eq:app-identity} explicitly. At $\beta=90^\circ$, both functions vanish in this lowest-order template. The nonzero high-$J$ states found numerically at exactly perpendicular finite drift arise from grain-specific higher-order angular structure that the template does not contain.

The weak-drift response of the complete facet torque gives $H(\Theta)=s_dD_\parallel\cos\Theta+O(s_d^2)$ and $F(\Theta)=\tfrac12s_dD_\parallel\sin\Theta+O(s_d^2)$. Therefore
\begin{align}
 H_B(\psi;\beta)&=s_dD_\parallel\cos\psi\cos\beta+O(s_d^2),\\
 F_B(\psi;\beta)&=\tfrac12s_dD_\parallel\sin\psi\cos\beta+O(s_d^2).
 \label{eq:app-weakB}
\end{align}
The common $\cos\beta$ factor makes the first-order magnetic-axis spin-up vanish at exact transverse drift. At $\beta=0$ the same weak-drift functions describe the drift-axis dynamics. They give $F'(0)/H(0)=1/2+O(s_d)$ when the axial response is nonzero.

\bibliographystyle{aasjournal}
\bibliography{references,met_shock_references}
\end{document}